\documentclass[prx,aps,10pt,twocolumn,superscriptaddress]{revtex4-1}
\usepackage{geometry}
\usepackage{graphicx}
\usepackage{amsfonts}
\usepackage{amsmath}
\usepackage{bm}
\usepackage{url}
\usepackage{color}
\usepackage{xfrac}
\usepackage{array,url,kantlipsum}

\definecolor{purple}{rgb}{0.5, 0.0, 0.8}
\definecolor{orange}{rgb}{0.9, 0.6, 0.0}

\begin{document}
\title{Self-assembly of complex structures in colloid-polymer mixtures} 

\author{Erdal C.~O\u{g}uz}
\email[]{erdaloguz@mail.tau.ac.il}
\affiliation{Institut f\"{u}r Theoretische Physik II, Heinrich-Heine Universit\"{a}t, Universit\"{a}tsstr. 1, 40225 D\"{u}sseldorf, Germany}
\affiliation{School of Mechanical Engineering and The Sackler Center for Computational Molecular and Materials Science, Tel Aviv University, Tel Aviv 6997801, Israel}
\author{Aleksandar Mijailovi\'c}
\affiliation{Institut f\"{u}r Theoretische Physik II, Heinrich-Heine Universit\"{a}t, Universit\"{a}tsstr. 1, 40225 D\"{u}sseldorf, Germany}
\affiliation{Institut f\"{u}r Theoretische Physik 1, Friedrich-Alexander-Universit\"{a}t Erlangen-N\"{u}rnberg, Staudtstr. 7, 91058 Erlangen, Germany}
\author{Michael Schmiedeberg}
\affiliation{Institut f\"{u}r Theoretische Physik 1, Friedrich-Alexander-Universit\"{a}t Erlangen-N\"{u}rnberg, Staudtstr. 7, 91058 Erlangen, Germany}

\date{\today}

\begin{abstract}
If particles interact according to isotropic pair potentials that favor multiple length scales, in principle 
a large variety of different complex structures can be achieved by self-assembly. We present, motivate, and discuss a conjecture for the occurrence of non-trivial 
(i.e., non-triangular) orderings based on newly-introduced enthalpy-like pair potentials, the capability of which we demonstrate for the specific example of 
colloid-polymer mixtures.  Upon examining the phase behavior of two-dimensional colloid-polymer mixtures, which can also be realized in experiments, we observe that 
non-trivial structures only occur in the vicinity of selected densities where triangular ordering is suppressed by the pair potential. Close to these densities, 
a large number of different phases self-assemble that correspond to tilings containing triangular, rhombic, square, hexagonal, and pentagonal tiles, and including some 
of the Archimedean tilings. We obtain the ground-state energies by minimizing the corresponding lattice sums with respect to particle positions in a 
unit cell as well as cell geometry and verify the occurrence of selected phases at finite temperatures by using Brownian Dynamics simulations. All reported phases should 
be accessible in experiments and, in addition, our work provides a manual on how to find the regions of non-trivial phases in parameter space for complex pair 
interactions in general.

\end{abstract}

\maketitle

\section{Introduction}

Self-assembly is the process by which the system constituents form large organized functional units via their mutual interactions without
any external influence. Since the pioneering work by Whitesides et al. \cite{Whi91} on molecular systems, self-assembly has been 
studied in great detail in a wide range of length scales ranging from atomic to macroscopic systems and throughout various scientific disciplines, 
including physics, chemistry, materials science, and biology \cite{Whi02,Whi02_pnas,Gro08,Leh02,Jen99,Jac04,Man03,Jac16,Kus69,Yve94,Bud95,Jul08,Dot14}.

In monodisperse systems that do not possess more than one characteristic length scale self-assembly usually occurs in the same way as for hard 
disks or spheres, i.e., the particles form a triangular phase in two or a fcc-crystal in three dimensions and at appropriate conditions.  
Beyond such simple systems, a plethora of self-assembled complex structures can be achieved in mixtures of different particle types such as 
in metallic alloys \cite{She84} or in binary colloidal suspensions \cite{Ass07,Ass08,Ant13}. Two further scenarios pave the way for high structural 
complexity in one-component systems by requiring \textit{i)} an interaction or shape anisotropy and \textit{ii)} a deliberate choice of an isotropic 
pair interaction with multiple length scales that affect the structure formation. Prototypical examples of the former are hard convex polyhedra packings 
\cite{Dam12} and patchy colloids \cite{Che11,Dop12,Bia14}, whereas concerning the latter inverse statistical-mechanical approaches have been undertaken
to investigate the self-assembly of non-triangular ground-state structures with engineered pair interaction potentials \cite{Rec05,Rec06,Bat08,reviewSal}.
More recently it has even been shown that complex ground-state structures can be stabilized by repulsive and convex pair interaction potentials which 
rule out the necessity for single- or multiple wells in this potential \cite{Coh09,Mar11,Jain13,Jain14,Pin16a,Pin16b}.
Unfortunately, the interactions obtained by the aforementioned inverse construction methods are most often artificial and 
in the consequence hard to realize in experiments.

A system that is well-studied in theory and simulation and in addition can be realized and investigated experimentally is a charge-stabilized 
colloid-polymer mixture which intrinsically involves multiple length scales as the colloids effectively attract each other close to contact due 
to depletion interactions while being repulsive on larger length scales because of screened electrostatic repulsions.
Experimental and theoretical studies have been performed to understand the nature of colloid-polymer mixtures \cite{Cas03,Dom12,Fen15,Pha02,Eck02,Man05,Lu08,
Zha13,Man14,Koh16,Str04,Taf10} some of 
which exhibit glassy states \cite{Pha02,Eck02}, gels \cite{Man05,Lu08,Zha13,Man14,Koh16}, and cluster formation \cite{Man14,Str04,Taf10}.
The phase behavior concerning the gas, liquid, and trivial solid phase has been investigated in \cite{Cal93,Dij99,Sch02,Poo02,Aar02,Roy05,For05,
Fle07,Fle08,Tui08,Gar16}. Moreover, the influence of many-body interactions on the phase behavior of such mixtures \cite{Jor03,Dij06} as well as confinement 
effects \cite{Cui05,Fry05,For06,Vin06} have been analyzed recently. 
However, to the best of our knowledge, detailed structural analysis of the crystal phase were not in the focus of previous studies, 
nor any non-trivial complex ordering have been reported with the exception of local ground-state clusters \cite{Taf10,Man14}. 

In this article, we determine the ground states that occur in a colloid-polymer mixture in two dimensions. The ground-state orderings are calculated 
by minimizing the energy for structures with one to six particles per unit cell. Aside from the trivial triangular phase, we observe square, rhombic, 
triangular, honeycomb, Kagome, and Archimedean tilings \cite{Kepler1619,Sch10,Ant11,Mil14} as well as further orderings that correspond to 
tilings with hexagons and even pentagons. 

In order to understand why non-trivial structures in monodisperse systems with isotropic interactions can occur in general, we introduce an 
enthalpy-like pair potential and conjecture that it can be used in order to identify the parameters such that triangular structures are suppressed. 
Our approach explains why and how even monotonic pair potentials can be used in order to self-assemble complex phases.

\section{System and Methods}

\subsection{Colloid-polymer mixtures}

When immersed in a solvent of relatively small-sized non-adsorbing polymer coils, the larger colloidal particles sense a short-ranged  \textit{depletion}
attraction at sufficiently high polymer concentrations. 
This attractive force arises due to an unbalanced osmotic pressure stemming from the depletion zone in the region between the colloids. 

The effective pair interaction potential between point-like colloids in the presence of the polymers is given by a screened Coulomb repulsion corresponding to 
a Yukawa-like interaction, and on short lengths by depletion attraction where we employ the AO-model \cite{Asa54,Vri76}. Therefore, the pair interaction potential is
\begin{equation}
\label{eq:ao_pairint} 
v(r_{ij})=
\begin{cases} 
V_0\dfrac{ \exp ( -\kappa r_{ij}) }{ \kappa r_{ij}} - W_0 \left[1-\dfrac{3r_{ij}}{2d} + \dfrac{r_{ij}^3}{2d^3}\right] \\ & \hspace{-1.3cm} \text{if}  \,\,\, r_{ij} \le d , \\
V_0\dfrac{ \exp ( -\kappa r_{ij}) }{ \kappa r_{ij}}    & \hspace{-1.3cm} \text{if}   \,\,\, r_{ij} > d ,
\end{cases}
\end{equation}
where  $r_{ij}$ is the separation distance between colloids $i$ and $j$, $\kappa$ denotes the inverse screening length, and $d$ the depletion length 
dictating the range of the attractive interaction and corresponding to the typical diameter of the polymers. The energy amplitude of the pure electrostatic 
Yukawa interaction is given by $V_0$, whereas the strength of the depletion potential is set by $W_0$. The crystalline phase diagrams can therefore 
be determined in three-dimensional space spanned by the reduced energy amplitude $V_0/W_0$, the reduced density $\sqrt{\rho} d$, and the reduced depletion 
length $\kappa d$ corresponding to  the ratio of depletion length divided by screening length. 

We primarily explore the ground state of our model colloid-polymer mixtures by determining the corresponding phase diagrams in the  
$(\sqrt{\rho} d, \kappa d)$-plane at fixed $V_0/W_0$. At zero-temperature, the optimal structures with $N$ particles are 
those that minimize the total internal energy 
\begin{equation}
\label{eq:ao_potnrg} 
U = \dfrac{1}{2} \sum_{\substack{i,j \\ i \ne j}}^{N} v(r_{ij})
\end{equation}
at a given reduced density, depletion length, and energy amplitude.
We use a direct lattice summation technique to determine $U$, and thus to predict the corresponding ground-state structures.  
In order to examine the stability of resulting structures at finite temperatures, we extend our studies to $T>0$ by means of Brownian Dynamics 
computer simulations. In the following, we provide details for both the lattice summation and the finite-temperature simulations used in this work.

\subsection{Lattice-sum calculations}

At each given density $\sqrt{\rho} d$, depletion length  $\kappa d$, and energy amplitude $V_0/W_0$, we have performed lattice sum 
minimizations for a set of candidates of crystalline lattices. As possible candidates, we consider two-dimensional crystals with a periodicity 
in both the spatial directions $x$ and $y$ whose primitive cell is a parallelogram containing $n$ particles. 
This parallelogram is spanned by the two lattice vectors $\mathbf{a} = a (1,0)$ and $\mathbf{b} = a \gamma (\cos\theta, \sin\theta)$, 
where $\gamma$ denotes the aspect ratio ($\gamma = |\mathbf{b}|/|\mathbf{a}| = b/a$), and $\theta$ is the angle between $\mathbf{a}$ and $\mathbf{b}$. 
We consider candidates with primitive cells comprising up to 6 particles, i.e., $n=1, \cdots, 6$, with no further restrictions. 

At prescribed parameters, the total potential energy per particle $u=U/N$ (cf.\ Eq.\ \ref{eq:ao_potnrg}) is minimized with respect to the particle 
coordinates of the basis, and the cell geometry. The latter includes a minimization process of $\gamma$ and $\theta$. 
To be specific, we implement the Nelder-Mead method (also known as downhill simplex method or amoeba method) to find the minimum of the 
energy functions without calculating their derivatives \cite{Nel65}. As this technique is a heuristic approach, and thus it may not always converge 
to the global minimum, we use at least $200$ and at most $1000$ different start configurations depending on the complexity of those functions.

\subsection{Enthalpy-like pair potential}
\label{sec:enthpairpot}

We introduce a new pair potential that corresponds to an enthalpy-like quantity $h(r_{ij})$. This enthalpy-like pair potential 
should be given such that the total enthalpy is
\begin{equation}
\label{totenth}
H=\frac{1}{2} \sum_{i,j, i \ne j}^{N} h(r_{ij})=U+pA,
\end{equation}
where $p$ is the macroscopic pressure and $A$ the total area of the system (in three dimensions the volume has to be used instead). 
The enthalpy-like pair potential therefore can be introduced as 
\begin{equation}
\label{eq:enth_pot} 
h (r) = v(r) + 2 \dfrac{p}{N}  a(r),
\end{equation}
where $a(r)$ is an effective surface area between any two particles at a relative distance $r$, which we refer to as 
the \textit{pair area} in the following. The pair area $a(r)$ has to follow from a subdivision of the total area $A$  as in
\begin{eqnarray}
\label{eq:Atot}
N A =  \sum_{\substack{i,j \\ i \neq j}}^N a(r_{ij}) .
\end{eqnarray}
Note that $a(r_{ij})$ shall be given such that for a fixed particle $i$, a summation over $j$ leads to $A$. Then, a summation over all $i$ and $j$ with 
$i\neq j$ yields a total area of $NA$.

In order to get an idea what the enthalpy-like pair potential can look like, we consider a particle $i$ and 
its $k$ neighbor shells. The remaining $N-1$ particles are distributed over these $k$ shells. We denote the $k$th shell of $i$,
to which the particle $j$ belongs, by the index $k_{ij}$, its thickness by $\delta_{k_{ij}}$, its relative position to the particle $i$
by $r_{k_{ij}}$, and the corresponding coordination number by $Z_{k_{ij}}$. Consequently, the pair area $a(r_{ij})$ can be written as 
the area of the $k$th shell of particle $i$ to which the particle $j$ belongs as  
\begin{equation}
 \label{eq:Apair} 
  a(r_{ij}) \simeq \dfrac{2\pi}{Z_{k_{ij}}} \delta_{k_{ij}} r_{k_{ij}},
\end{equation}
where $1/Z_{k_{ij}}$  compensates the counting  of multiple particles in the same shell so that each shell contributes only once 
to the sum in Eq.\ \ref{eq:Atot}. %The both sides are strictly equal if $r_k$ points to the exact center of the corresponding shell.
While the coordination number can be easily obtained from the theta series of lattices, the most common ones of which 
being tabulated in \cite{Slo87,Con99}, the choice of thickness $\delta_k$ is rather not unique as the space can be subdivided differently 
into circular non-overlapping rings under the constraint that each of which contain solely one neighbor shell. 
In general, the pair area can be written as $a(r) = f(r) r$, where $f(r)$ denotes the thickness $\delta$ of a shell at a distance $r$. 
On average, $f(r)$ is a decaying function with distance as the neighbor shells become closer to each other for large $k$. A canonical 
choice for the thickness would be $\delta_k = (r_{k+1} - r_{k-1})/2$, where the shell positions $r_k$ could be gained 
from the radial distribution function of the structure.
However, within the range of the first few neighbor shells, $\delta_k$ can be chosen to be a constant. 
The value of $Z_k$ behaves similarly in the same range. For example, in case of the triangular lattice, the first three shells possess $Z_k = 6$ 
particles ($k=1,2,3$). Usually, only the closer neighbors will play the predominant role for the stability. If this is the case, we finally find the 
following approximate functional form for the pair area $a(r)$ for distances corresponding to a few inner neighbor shells:
\begin{equation}
 \label{eq:arapprox}
  a(r) \approx C(\rho, Z, \tilde{\delta}) \kappa r,
\end{equation}
where the density-dependent prefactor $C(\rho, Z, \tilde{\delta})$ incorporates a constant coordination number $Z$ and a constant reduced thickness 
$\tilde{\delta} = \delta / a_1$  for the first few shells with $a_1$ being the nearest-neighbor distance, but approximately no dependence on $r$.

\subsection{Brownian Dynamics computer simulations}

To study the validity of our theoretical ground-state predictions at finite but relatively low temperatures, 
we employ Brownian Dynamics computer simulations in the $NVT$-ensemble by solving the Langevin equation for an overdamped system. 
The position $\mathbf{r}_i$ of particle $i$ undergoing Brownian motion after a
time step $\delta t$ is
\begin{equation}
\label{eq:eot} 
\mathbf{r}_i (t + \delta t) = \mathbf{r}_i (t) + \dfrac{D}{k_BT} \mathbf{F}_i (t) \delta t + \delta \mathbf{W}_i ,
\end{equation}
where $D$ denotes the free diffusion coefficient, $k_BT$ the thermal energy, and $\mathbf{F}_i$ is the total conservative force acting on particle
$i$ and stemming from the pair interaction $v$ in Eq.\ \ref{eq:ao_pairint}.
The random displacement $\delta \mathbf{W}_i$ is sampled from a Gaussian distribution with zero mean and variance $2D\delta t$ (for each Cartesian 
component) fixed by the fluctuation-dissipation relation.
The time step is chosen as $\delta t = 10^{-5} \tau$ where  $\tau = 1 / (\kappa^2 D_0)$ is used as the unit of the time. 
We run simulations for up to $10^5 \tau$, starting from a random distribution, a triangular or a square lattice of $N = 2000$ particles in a rectangular 
simulation box under periodic boundary conditions. All three runs yield the same final configurations at predetermined 
density and depletion length, suggesting the thermodynamical stability of our results rather than possible metastable configurations.

It is noteworthy that more sophisticated simulation algorithms making use of non-rectangular simulation boxes \cite{Gra12} 
might yield stable phases which we are not able to capture here in our simulations. This being said, however, we do not expect a radical change in the morphology of the 
phase diagram; In soft systems, the coexistence between two main phases will most likely suppress the occurrence of subtle phases with free-energies 
relatively close to each other and to the aforementioned main phases. Moreover, we want to emphasize that our main goal lies in determining the 
ground-state of colloid-polymer mixtures, where we consider non-rectangular boxes, and comparing the results to the predictions of our enthalpy-based theory, 
where simulations shall only serve as an additional supportive data to strengthen our findings.

In the following, we present the results from both the zero-temperature lattice-sum minimizations and the finite-temperature simulations.

\section{Results}

Before delving into the phase diagram for the given example of colloid-polymer mixtures in detail, i.e., determining each single phase structure 
at prescribed density and depletion length, we first explore the conditions that lead to the formation of non-trivial structures.

\subsection{Enthalpy-like pair potential and occurrence of non-triangular phases}

\begin{figure}[h!]
 \centering
 \includegraphics[height=6.0cm]{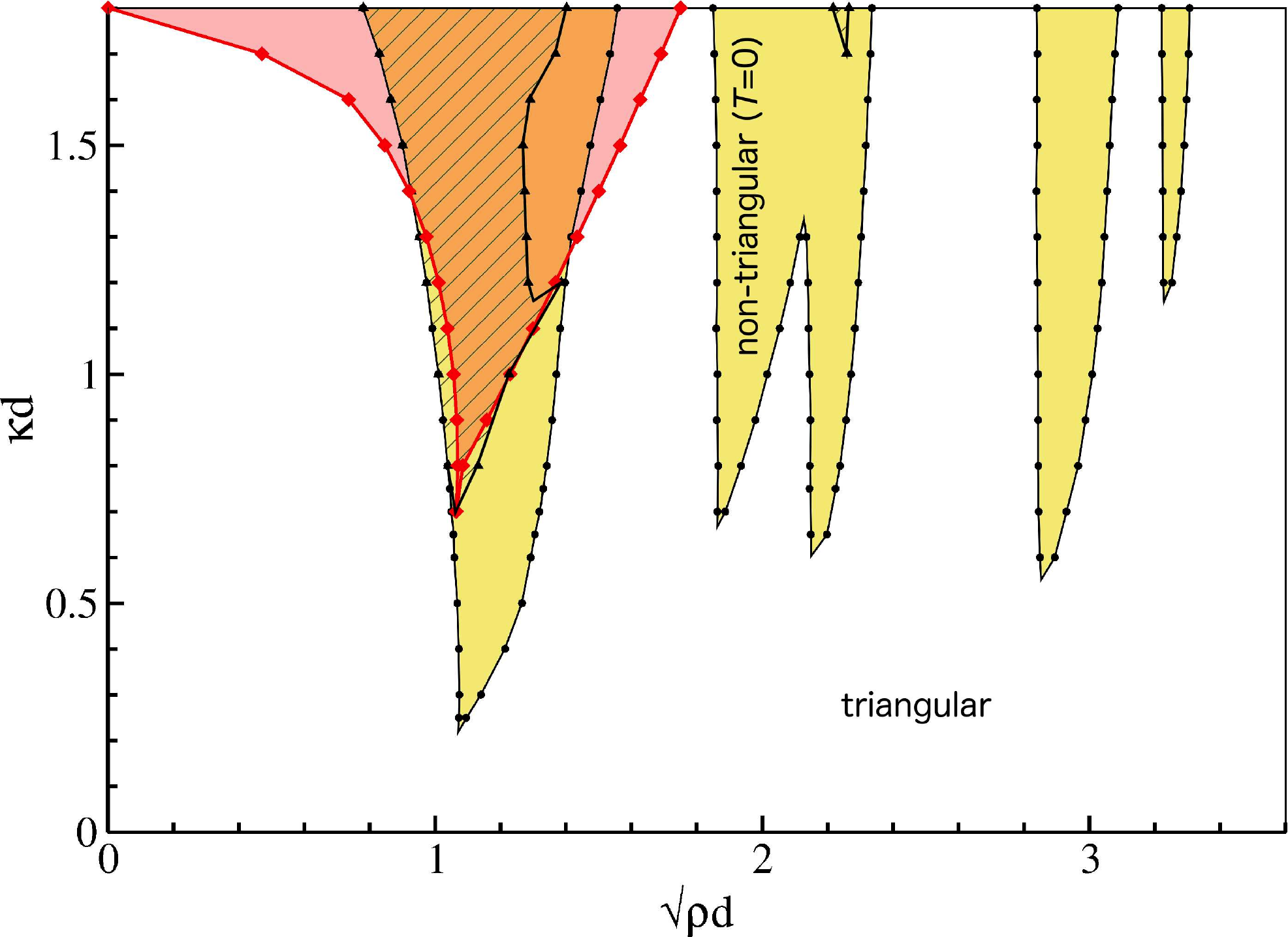}
 \caption{Ground-state phase diagram of colloid-polymer mixtures as a function of the reduced density $\sqrt{\rho} d$ and the reduced 
 	  depletion length $\kappa d$. The majority of the phase space (white area) is governed by the triangular phase, whereas non-triangular phases 
	  can be stabilized around certain values of the reduced density and along the reduced depletion length. 
	  These non-triangular stability modes as determined by lattice-sum minimizations are shown by yellow regions. 
	  The actual determined phase points at the boundary are indicated by black dots. The red region indicates the theoretically predicted stability 
	  zone where complex non-triangular structures with two different length scales  that are sufficiently distinct from the triangular length scale 
	  are expected to become stable. The hatched non-triangular subdomain comprises such stable complex phases as obtained by lattice-sum minimizations 
	  where the closest-neighbor distance deviates at least fifteen percent from the triangular lattice constant.}			 
 \label{fig:phdiag1}
\end{figure}

As shown in  Fig.\ \ref{fig:phdiag1}, for colloid-polymer mixtures, the non-triangular crystalline phases only occur in special 
parameter regions. These regions are marked in Fig.\ \ref{fig:phdiag1} in yellow for ground states as determined by minimizing the interaction energy. 
The red area indicates the phase space within which complex non-triangular structures are expected to become stable according to our enthalpy-based theory. 
On the one hand, the lattice-sum minimization process yields precise results (within the numerical accuracy), and as such, 
the corresponding ground-state phase diagram is accurate. On the other hand, our theory delivers a good estimate for the 
occurrence of complex non-triangular phases if their stability is dominated by the closest neighbors.
Among these non-triangular phases are the complex 
structures with two length scales that are sufficiently distinct from the triangular length scale such as in case of the honeycomb lattice.

The differences between zero and non-zero temperature are discussed in the next subsection, and the non-triangular structures are analyzed in detail in 
Sec. \ref{sec:non-tri}. In the following, we first explore the stability of non-triangular phases in detail, and we describe our novel theoretical
approach for the occurrence of complex non-triangular orderings. 

The triangular lattice can be stabilized by various types of two-body interactions, among which are those that are purely repulsive and 
that involve one simple length scale. For example, a screened Coulomb potential (corresponding to a Yukawa interaction), which is widely used 
to model charge-stabilized colloidal suspensions, leads to triangular ordering. As opposed to such interactions, the presence of multiple length scales in the 
pair interaction potential might yield two-dimensional crystals other than the triangular lattice \cite{reviewSal}. For example, the interparticle forces in 
the colloid-polymer mixture are dictated by two different length scales that define the ranges of the attraction and the repulsion.

\begin{figure}
 \centering
 \includegraphics[height=5.6cm]{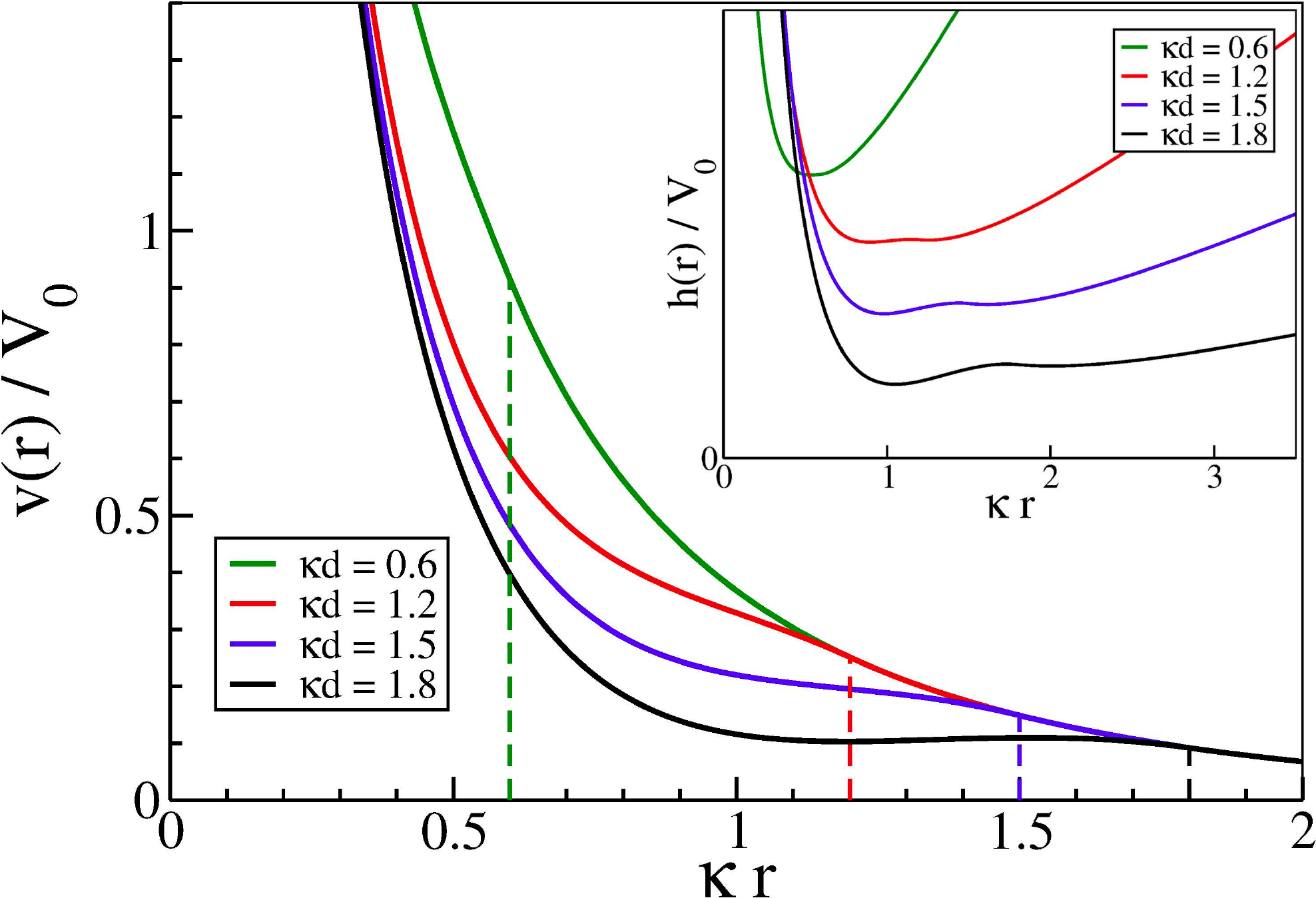}
 \caption{Characteristic examples of pair interaction potentials $v(r)$ as given by Eq.\ \ref{eq:ao_pairint}. Potentials are shown for three
    	  different reduced depletion lengths ($\kappa d = 0.6, 1.2, 1.5, 1.8$), where the depletion lengths are indicated by 
	  the dashed lines. The inset illustrates the enthalpy-like pair interaction for $\kappa d = 0.6, 1.2, 1.5, 1.8$ at 
	  $\sqrt{\rho}d = 1.1$ as defined in Eq.\ \ref{eq:enth_pot}. 
	  Although $v$ for $\kappa d = 1.2$ and $\kappa d = 1.5$ does not possess any local maxima and minima, the corresponding $h$ features a concave
	  region separating two minima with a local maximum, yielding the stability of complex structures with multiple length scales, 
	  whereas $h$ for $\kappa d = 0.6$  lacks the existence of any concave region. Note that, in the inset, the enthalpy-like pair interactions are arbitrarily
	  shifted along the $y$-axis for the sake of clarity.}
 \label{fig:pairpots}
\end{figure}

In order to understand the nature of non-triangular stability modes, we first consider the explicit form of the pair interaction in 
Eq. (\ref{eq:ao_pairint}), characteristic examples of which are shown in Fig.\ \ref{fig:pairpots} at different reduced depletion lengths $\kappa d$ 
and at $V_0/W_0 = 1$.  For large $\kappa d = 1.8$, a single well occurs with a clear local maximum slightly below the reduced depletion length 
(the positions of which are indicated by dashed lines). We find that the non-triangular structures occur at reduced densities where 
the depletion length $d$ approximately corresponds to the $k$th next-neighbor distance $a_k$ of the triangular lattice with $k=1,\cdots,5$, 
cf.\ Fig.\ \ref{fig:phdiag1}.
These distances are $a_1, a_2 = \sqrt{3} a_1 $, $a_3 = 2 a_1$, $a_4 = \sqrt{7} a_1$, and $a_5 = 3 a_1$ with $a_1 \approx 1/\sqrt{\rho}$.  
In other words; the triangular lattice is suppressed at densities where its interparticle distances roughly correspond to the depletion length of the system 
and thus being in the close vicinity of the maximum of the potential yielding an increase in the energy of the triangular lattice. 
The system will therefore possess lower potential energy for structurally more complex crystals with separated length scales or for rhombic or square lattice.

Non-triangular phases are still present at low reduced depletion lengths where the pair interaction potential does not possess a local maximum, e.g., 
for $\kappa d = 1.2$ as shown by the red curve in Fig.\ \ref{fig:pairpots}.  Therefore, the triangular order can even be suppressed without a maximum in the 
pair interaction potential. Note that the presence of concavity is also not a necessary condition to stabilize non-triangular structures. Recently developed 
inverse statistical-mechanical methods have been indeed used to engineer purely repulsive and convex interaction potentials that yield complex 
ground-state crystals such as honeycomb lattice \cite{Mar11,Pin16a,Pin16b,Jain13,Jain14}. In the following, we want to develop a general method to  
predict the occurrence of complex non-triangular phases for a given pair interaction based on the enthalpy-like pair potential introduced 
in Sec. \ref{sec:enthpairpot}.

In the following, we analyze the conditions under which $h(r)$ might exhibit two minima. Note that, unlike the pair interaction potential $v(r)$, 
the explicit form of $h(r)$ depends on the density via the macroscopic pressure $p$, cf.\ Eq.\ \ref{eq:enth_pot}. 
The pressure $p$ is given by $p = - \partial U / \partial A$ where in the vicinity of a given particle $i$ we can use 
$dA_j=d\left(\pi r_{ij}^2\right)=2\pi r_{ij}dr_{ij}$ as the change in the area and the energy per particle 
$u = U/N =\sum_{i,j,i \neq j}v(r_{ij})/2N \approx \sum_{j}v(r_{ij})$ for $T=0$ such that 
$p/N\approx -\sum_{j} \frac{1}{2\pi r_{ij}}\partial v(r_{ij})/\partial r_{ij}$ in the ground state.
The prefactor $C(\rho, Z, \tilde{\delta}) = 2\pi \tilde{\delta} a_1 /\kappa Z = \sqrt{8} \pi \tilde{\delta} / ( \sqrt{\sqrt{3}} \kappa Z \sqrt{\rho})$ 
is obtained from Eq.\ \ref{eq:Apair} which assumes areas of circular rings of a given thickness 
$\delta$ and with radii given by distances that occur in a triangular lattice.  
At each density and depletion length, we calculate $C(\rho, Z, \tilde{\delta})$ and $p/N$ for our reference system, i.e., the triangular lattice, 
and determine $h(r)=v(r) + 2 \dfrac{p}{N} C(\rho, Z, \tilde{\delta})\kappa r $.

In Fig. \ref{fig:phdiag1} we illustrate the phase space where $h$ possesses two local minima with a concave region between them by the red area 
(using $\tilde{\delta} \approx 0.4$). 
Within and only within the red area in Fig.\ \ref{fig:phdiag1}, we expect the stability of complex non-triangular phases if stability is dominated by the 
closest neighbors. If, however, the closest-neighbor distance in a non-triangular crystal deviates only slightly from the triangular lattice constant at 
the same density, it is then upon further neighbors whether the triangular or the non-triangular one becomes stable. For example, 
some regular phases like the square phase or rhombic phases that are morphologically close to the square phase are dominated by one 
nearest-neighbor length scale exactly as the triangular phase and therefore whether a triangular, a square or such a rhombic phase is stable not only depends 
on the closest neighbors. As a consequence, our theory can correctly predict a subdomain of the non-triangular stability
region, and thus the red area shown in Fig.\ \ref{fig:phdiag1} differs from the non-triangular yellow area obtained by ground-state calculations, if, 
e.g., a square phase occurs.

Attention must be paid when interpreting the red stability region of non-triangular complex phases as predicted by our theory, see Fig.\ \ref{fig:phdiag1}, 
where $h(r)$ possesses a concave region separating two local minima. The triangular phase within this area only becomes 
unstable if one of the particles of the triangular structure is actually located in the concave region of $h(r)$.
For large $\kappa d$, however, concave regions might occur in places where there is no particle in a triangular phase such that the triangular phase 
remains stable. As a consequence, the red area extends outside of the exact stability zone of the non-triangular phases obtained by ground-state calculations.

As opposed to the situation above, our theory captures very well the occurrence of some regular (e.g., elongated rhombic phases as shown in Fig.\ \ref{fig:pd_simple}b)
and complex phases (e.g., honeycomb, cluster phases, etc.) that are dominated by two nearest neighbors with
distances sufficiently distinct from the triangular lattice constant. Recall that if nearest-neighbor distances are similar to the 
triangular length scale, the stability of the corresponding phase over the stability of the triangular phase is up to farther particles, 
and thus our approach may not necessarily yield a solid prediction of those phases.  
Consequently, the phases with two length scales fairly deviating from the triangular one can only occur within the red area. 
As an example; the hatched area in Fig. \ref{fig:phdiag1} indicates a subspace of non-triangular phases with stable regular and complex phases whose 
nearest-neighbor distances differ at least 15 per cent of the triangular lattice constant at the same density. 
Note that the choice of such a distance cutoff is rather not unique. However, we observe the same qualitative picture for different cutoffs above 10 percent.
%Furthermore, up to some small deviations due to the employed approximations, we find that the triangular phase is only found outside the red area.

To demonstrate the explicit form of the enthalpy-like pair potential $h(r)$ in contrast to the pair potential $v(r)$, 
we plot $h(r)$ in the inset of Fig.\ \ref{fig:pairpots} for $\kappa d = 0.6, 1.2, 1.5, 1.8$ with 
$\tilde{p}\equiv 2pC/NV_0 \approx 2.63, 0.41, 0.20, 0.10$, respectively. These reduced pressure 
values are obtained at $\sqrt{\rho} d =1.1$. Three of them, namely $\kappa d = 1.2, 1.5, 1.8$ show a clear local maximum alongside the two minima, 
whereas $h$ for $\kappa d =0.6$ lacks completely such a situation for any density. A comparison with Fig.\ \ref{fig:phdiag1} confirms our conjecture that 
if $h(r)$ has no concave region such as for $\kappa d = 0.6$, then we do not expect the stability of complex phases with multiple length scales that are 
fairly distinct from the triangular length scale.

Note that concerning the whole system, we are in an $NVT$-ensemble such that employing an enthalpy-like quantity is obviously unusual. 
This being said, we try to find a local criterion that can be used to predict the global structure from local properties. 
The local structure of an infinite thermodynamic system as given by a particle and its neighbors does not necessarily possess the same 
thermodynamic properties as the global structure. However, for a given reference structure, 
the local arrangement clearly depends on the pressure such that locally a free enthalpy-like quantity controls the order. 
Accordingly, at zero temperature an enthalpy-like function as we introduced here can be used to primarily predict the occurrence of structures 
with multiple length scales.

The global order is still reflected by the pressure as well as the functional form $a(r)$ used in order to determine $h(r)$. 
As a consequence in principle $h(r)$ has to be determined for different candidate structures, in order to check whether the corresponding 
candidate structure minimizes the total enthalpy $H$ as given in Eq.\ \ref{totenth}. 
However, since the functional form in  Eq. (\ref{eq:arapprox}) is a good approximation to many lattices, 
the enthalpy-like function obtained by this choice for $a(r)$ can be used to check whether the triangular order is stable or not.
Furthermore, it is noteworthy that our criterion for non-triangular order does not take three-body or other multi-body interactions into account.

To summarize the results of this section, by determining the enthalpy-like pair potential $h(r)$ for triangular lattices we can predict the parameters 
where triangular order or other symmetries dominated by only one length scale can become unstable, namely if $h(r)$ possesses a concave part. 
Furthermore, we know that complex phases that are dominated by two 
distances between nearest neighbors can only occur if such a concave part exists. 
Note that instead of complex phase with two or more length scales in principle also a coexistence between phases with different length scales might occur.
By refining our method by including not only the closer neighbor structure, it should be able to improve the predictions. 
However, then the determination of $h(r)$ would become much more complicated and the benefit of a criterion that can be checked very easily for many parameters 
would be lost. 

In Sec. \ref{sec:non-tri} we analyze in detail which phases can occur for the example of a colloid-polymer mixture if the triangular phase is not stable. 
In the next subsection, we first discuss the differences between the calculated ground-state results and our simulation results for finite temperature.

\subsection{Zero- vs. finite-temperature results}

\begin{figure}
 \centering
 \includegraphics[height=5.8cm]{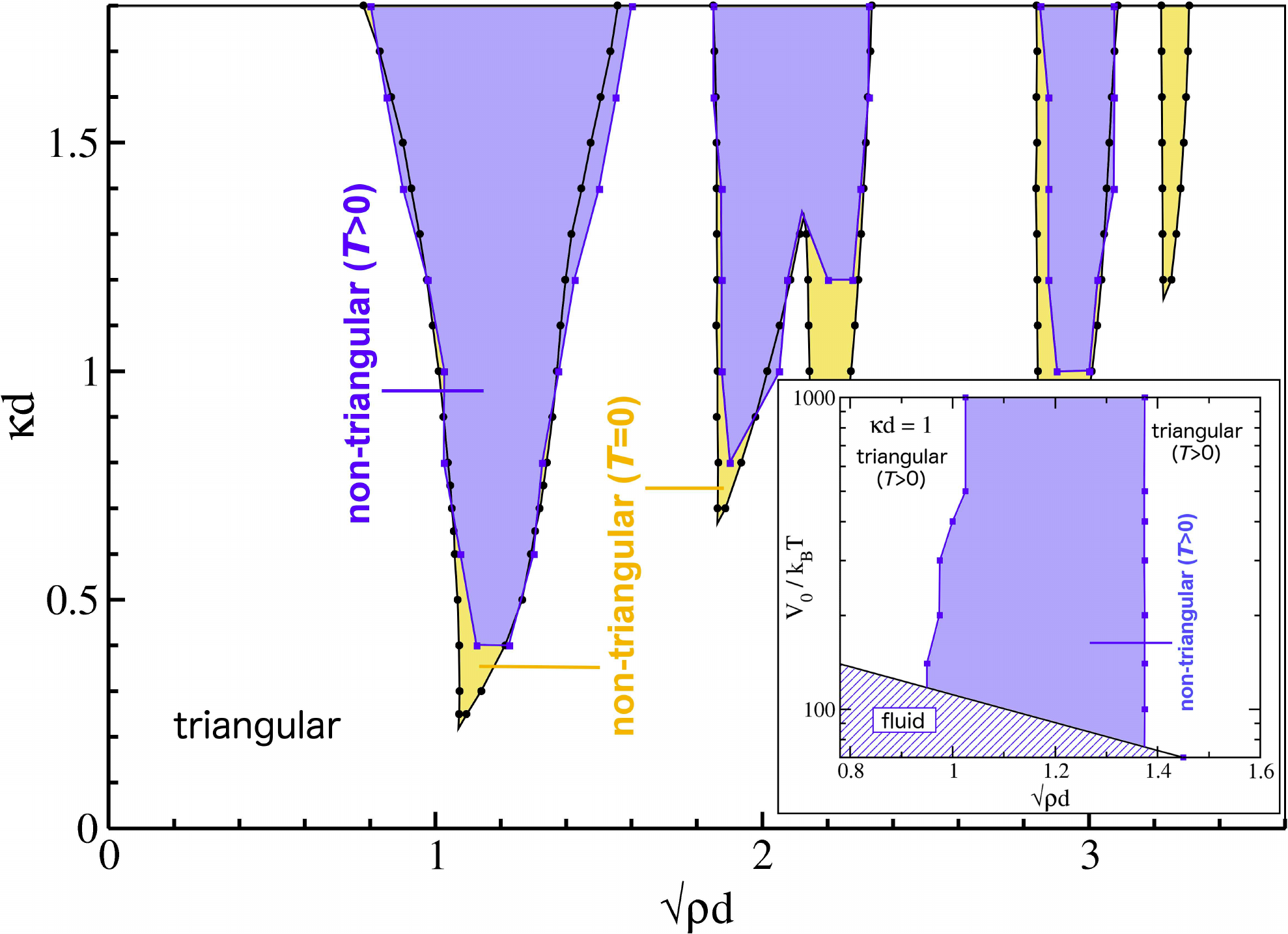}
 \caption{Comparison of zero- and finite-temperature phase diagrams of colloid-polymer mixtures as a function of the reduced density $\sqrt{\rho} d$ and 
 	  the reduced depletion length $\kappa d$.
	  The non-triangular stability modes as determined by lattice-sum minimizations and Brownian Dynamics simulations are shown 
	  by yellow (at $T=0$) and blue (at $T>0$) regions, respectively. The actual determined phase points at the boundary are indicated by black dots 
	  and blue squares. The inset shows the solid- and fluid-state phase diagram as a function of the reduced density and the reduced inverse 
	  temperature at fixed $\kappa d = 1$. Details are explained in the text. }			 
 \label{fig:phdiag2}
\end{figure}
We have investigated the phase diagram of colloid-polymer mixtures at finite temperatures by means of Brownian dynamics computer simulations. 
Particularly, we consider a temperature such that $V_0/k_BT = 1000$ and we fix the energy amplitudes to $V_0/W_0 = 1$.  
We reveal the stability of non-triangular crystals in the blue regions shown in Fig.\ \ref{fig:phdiag2}. The similarity to
the non-triangular stability regimes at $T=0$ (shown in yellow in Fig.\ \ref{fig:phdiag2}) is striking. Specifically, in both cases, 
the majority of the parameter space in the $(\sqrt{\rho} d, \kappa d)$-plane 
is governed by the triangular crystal as shown by the white region in Fig.\ \ref{fig:phdiag2}. Non-triangular phases only occur in the vicinity of 
specific densities and for sufficiently large values of $\kappa d$.

Moreover, the inset of Fig.\ \ref{fig:phdiag2} provides the evolution of the non-triangular solid phase space at $\kappa d = 1$ as a function 
of the inverse reduced temperature where melting is observed upon rising the temperature. Note also that at finite temperatures, the system 
melts as $\sqrt{\rho} d \to 0$. The melting is, however, out of the scope of this paper, and therefore not analyzed further here.

Another striking feature in simulations is that we observe phases of almost all of the different phase categories as enlisted and explained in the next 
subsection. As the main goal of this manuscript is to provide a postulate based on which complex crystals can be predicted and compare the theoretical 
predictions to numerical results at $T=0$, simulations are just supposed to serve as additional data to strengthen the ground-state results. Therefore, 
we choose to show only a few characteristic simulation snapshots rather than describing the finite-temperature phase diagram thoroughly.

\subsection{Detailed phase diagram of a colloid-polymer mixture}
\label{sec:non-tri}
\begin{figure}[h!]
 \centering
 \includegraphics[height=5.8cm]{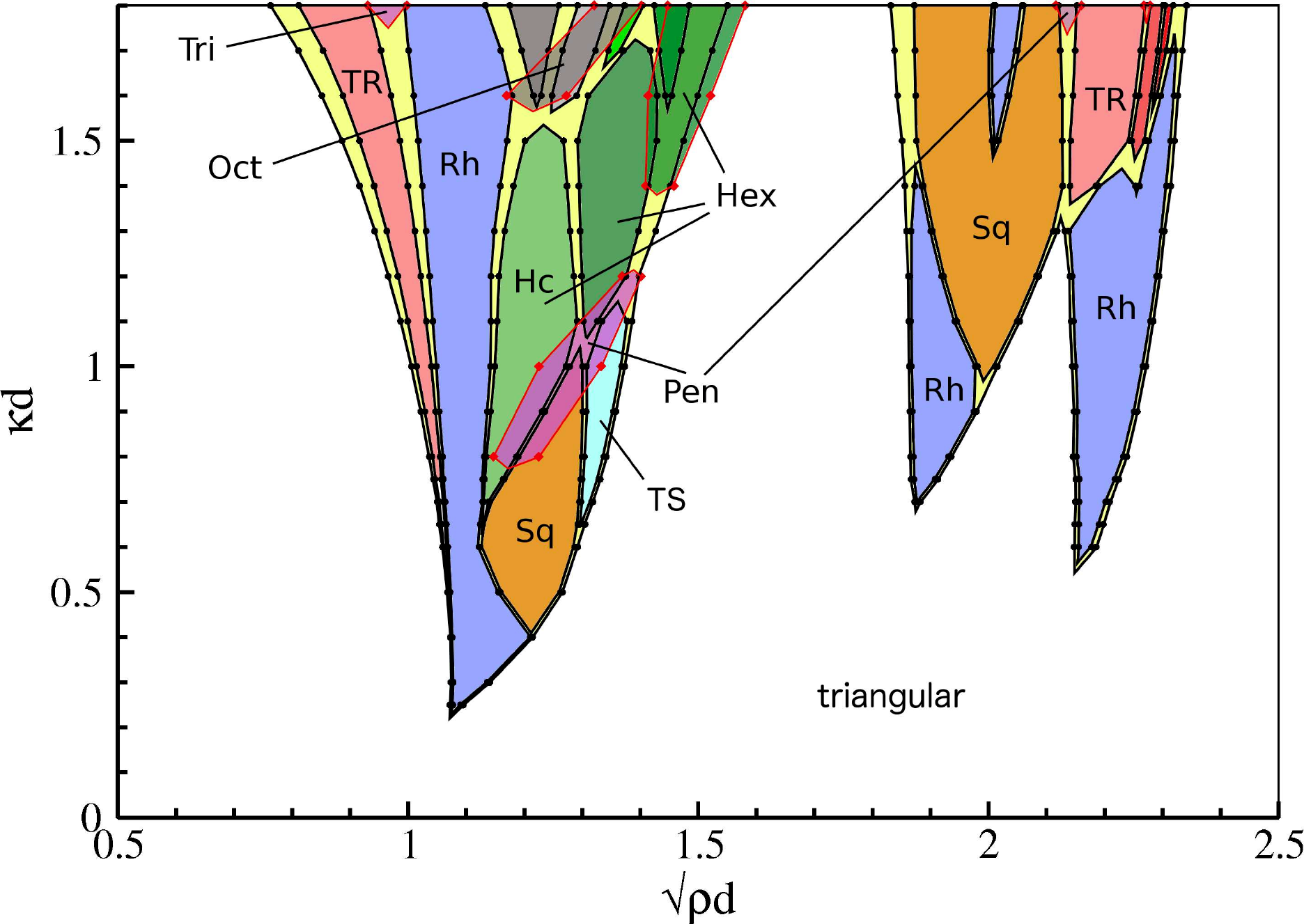}
 \caption{Detailed zero-temperature phase diagram of a colloid-polymer mixture as a function of the reduced density $\sqrt{\rho} d$ and the reduced
          depletion length $\kappa d$. The large white zone indicates the stability of the triangular phase, whereas other colors have been used to demonstrate 
 	  the non-triangular stability regimes that include a variety of phase structures, each colored differently and labeled as follows: Sq: square phase, 
	  Rh: rhombic phase, Hex: hexagon-based structures including the honeycomb (Hc) lattice, Oct: octagon-based structures, Tri: trimers, 
	  TR: triangle-rectangular structures, TS: triangle-square structures, and Pen: pentagon-based structures. 
	  The black dots are the actual computed phase points at the boundary of non-triangular phases. 
	  The yellow areas illustrate the coexistence between the neighboring phases. The phases are shown for up to $4$ particles per unit cell. 
	  The regions where we find additional phases with 5 or 6 particles per unit cell are marked by transparent colors. 
	  While outside of these regions we do not expect the occurrence of phases with even more particles per unit cell, within the transparent regions 
	  more complex phases might be stable.}
 \label{fig:pd_detail}
\end{figure}

In the following, we explore the ground-state phase diagram of a colloid-polymer mixture in detail. The stability of zero-temperature crystalline phases are
shown in Fig.\ \ref{fig:pd_detail} for $\sqrt{\rho} d \le 2.5$ and $\kappa d \le 1.8$.%, i.e., for the first three  
%non-triangular stability modes at $T=0$. We further compare these results to our findings at $T>0$. 
%
In Fig. \ref{fig:pd_detail} we show the detailed phase behavior for the first non-triangular regions with lowest density.
The phase diagram reveals a large structural diversity. The white region indicates the stability regime of the triangular lattice as before, 
whereas the colored areas demonstrate the occurrence of stable non-triangular phases of different symmetry and complexity. 
Here, we obtain simple phases with $n=1$ particle per unit cell such as rhombic (Rh) and square (Sq) lattices as shown by blue and orange 
regions in Fig.\ \ref{fig:pd_detail} as well as more complex structures with $n \ge 2$. The latter possess a richer diversity and can be grouped into 
hexagon-based (Hex, green regions) structures including the honeycomb (Hc) lattice, octagon-based (Oct, gray regions) structures, trimers
(Tri, purple), triangle-rectangular (TR, red regions), and triangle-square (TS, turquoise region) structures, and finally 
the pentagon-based (Pen, purple transparent regions) structures as indicated by different colors and labeled accordingly in Fig.\ \ref{fig:pd_detail}. 
Structural details are provided below.

For the sake of completeness, we further investigated the phase coexistence in our system: We have implemented the common tangent construction 
(Maxwell construction) 
and have determined the corresponding coexistence regimes between two neighboring phases, 
which we indicate by the yellow areas in the phase diagram in Fig.\ \ref{fig:pd_detail}. As expected, the coexistence turns out to be relatively small at 
zero-temperature as compared to pure one-phase stability regimes. 

The phases in Fig.~\ref{fig:pd_detail} are shown for up to $4$ particles per unit cell with solid colors. In addition, the regions where 5 or 6 
particles per unit cell lead to new phases are marked by transparent colors encircled by the red lines. While outside of these regions we do not have any 
indication for the occurrence of phases with even more particles per unit cell, within the transparent regions more complex phases might be expected to be stable.

\subsubsection{Triangular (Tr), rhombic (Rh), and square (Sq) phases}

\begin{figure}[h!]
 \centering
 \includegraphics[height=14.0cm]{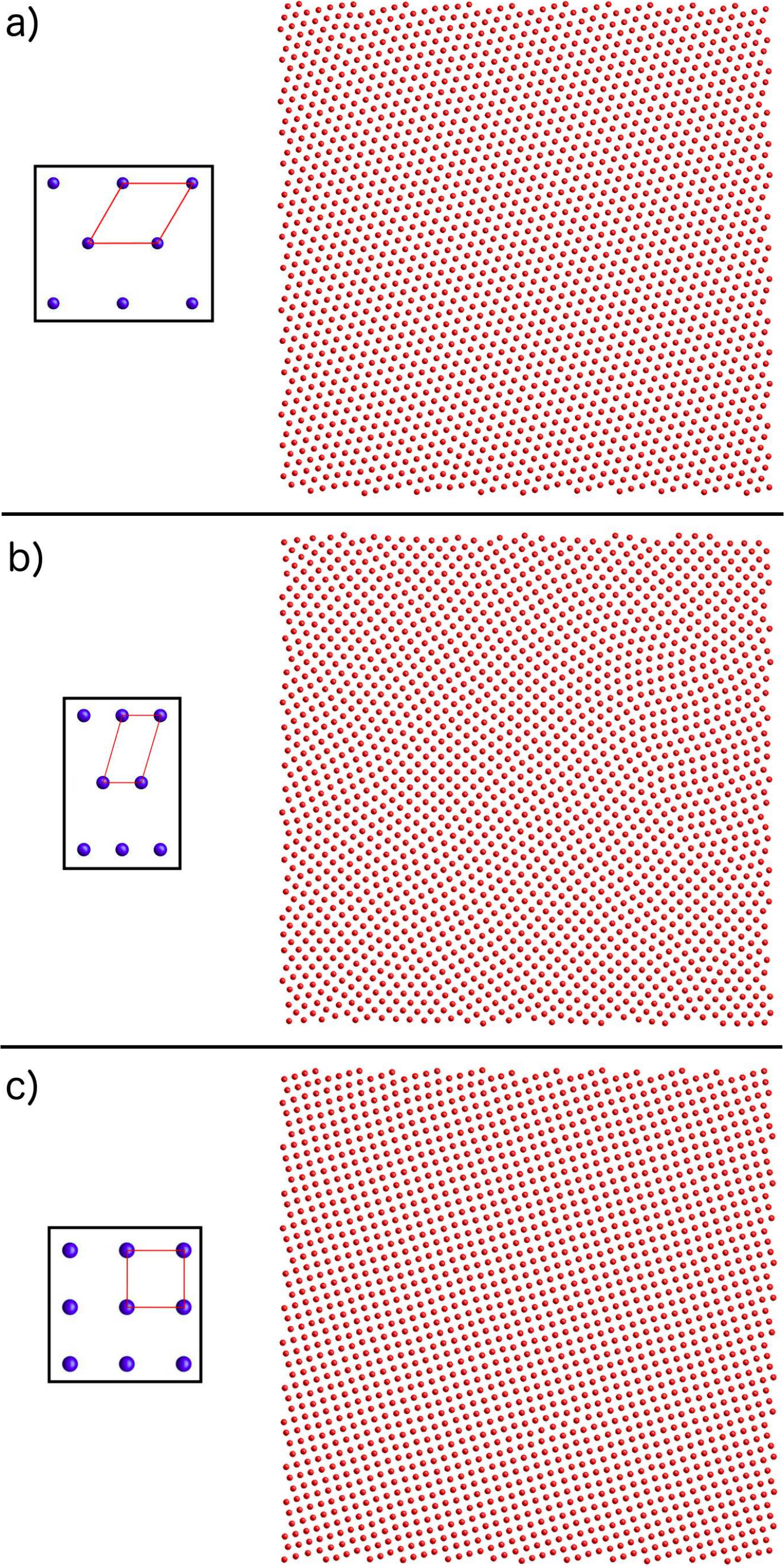}
 \caption{Triangular (a), rhombic (b), and square (c) phases at $T=0$ (left panels) as obtained by lattice-sum minimizations and 
 	  $T>0$ (right panels) as obtained by Brownian Dynamics simulations. The red lines on the left panels depict the unit cells of the structures. 
	  The simulations contain $N=2000$ particles.}
 \label{fig:pd_simple}
\end{figure}

The ground-state phase diagram exhibits three simple crystalline phases, namely the triangular, rhombic, and square phases that   
are each shown in the left panels of Figs.\ \ref{fig:pd_simple}a, b, and c, respectively. 
The red lines serve as a guide to the eye showing the trivial unit cells in each structure. 
Characteristic snapshots of simple phases from finite-temperature BD simulations at $V_0/k_BT = 1000$ are displayed in the right 
panels of Figs.\ \ref{fig:pd_simple}a, b, and c. 

We further identify plenty of complex structures containing at least two particles per unit cell. 
To achieve a clear overview for the reader, we choose to present them in groups according to the main repeating structural unit of
each phase. Each group is uniquely color-coded in Fig.\ \ref{fig:pd_detail} as we use different colors and its shades for different groups. 
In the following, we list these groups alongside with structural images, and for clarity, we further regard the resulting structures as 
tilings, and we illustrate their prototiles.

\subsubsection{Hexagon-based structures (Hex)}

\begin{figure}[h!]
 \centering
 \includegraphics[height=5.0cm]{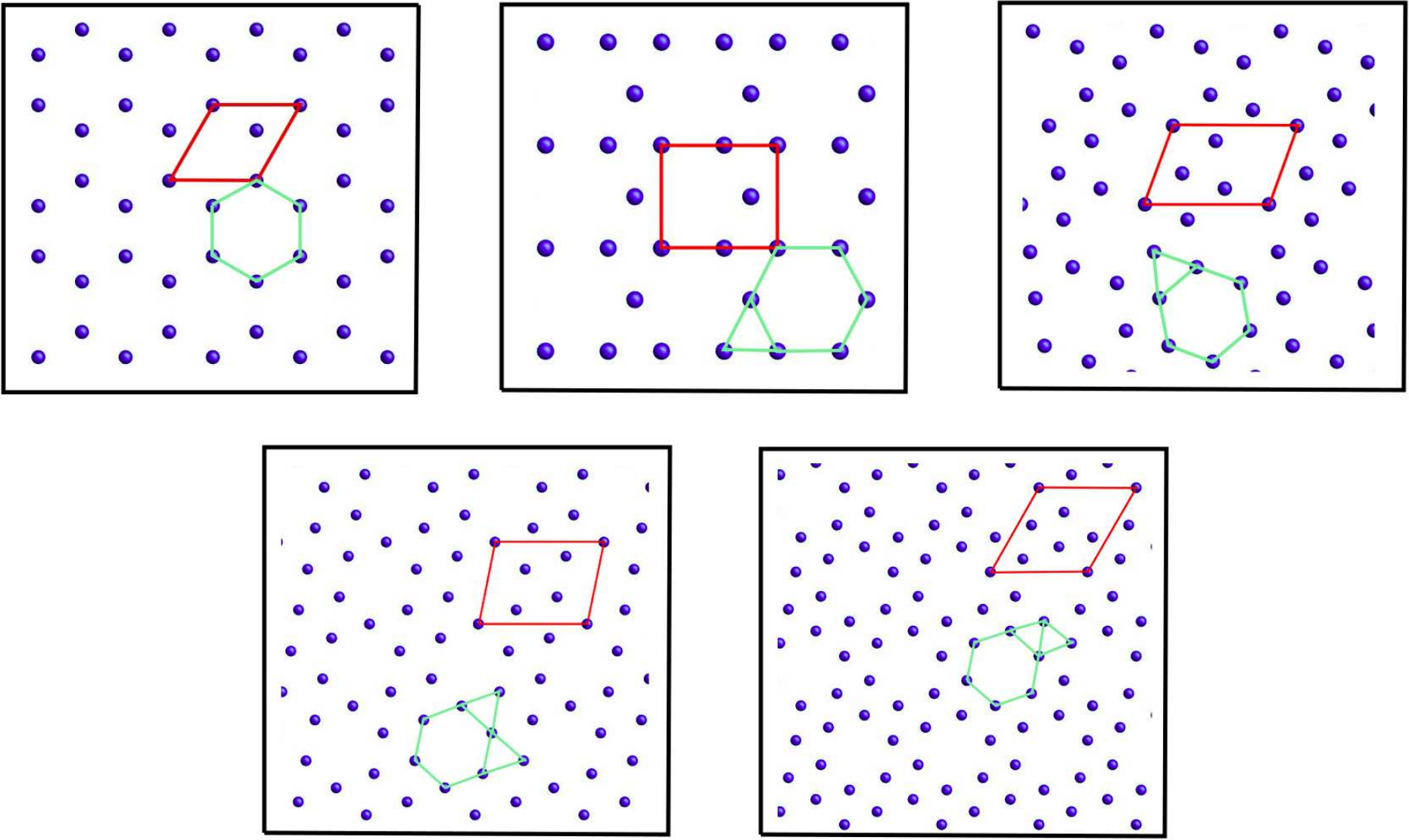}
 \caption{Schematic illustration of five stable hexagon-based structures with regular hexagons as their main repeating units 
 	  at $T=0$ as obtained by lattice-sum minimizations. 
	  Different prototiles of the corresponding tilings are indicated by green lines. These are either one hexagon or an hexagon with one or two 
	  different triangles. Note that the upper left structure corresponds to the honeycomb lattice. Red lines emphasize the unit cells with $n=2,\cdots,6$.}
 \label{fig:Hex}
\end{figure}

Hexagon-based structures appear in the first non-triangular region, albeit comprising regular hexagons as their main periodically repeating unit. 
For the sake of clarity, we provide a further point of view on the structures in the remainder of this paper: 
In addition to the unit cells that are marked red, we show the characteristic prototiles of the corresponding tilings by green lines in each figure. 
In this case, the prototiles are either a hexagon or a hexagon together with one or two distinct triangles as indicated by the green lines in Fig.\ \ref{fig:Hex}. 
One of these structures corresponds to the well-known honeycomb lattice (upper left in Fig.\ \ref{fig:Hex}), whereas the other four consist of hexagons and 
triangles each in different stoichiometric ratios. This group is indicated by shades of green in Fig.\ \ref{fig:pd_detail} including the
green transparent region with $n=5,6$. The latter are shown in the lower panel of Fig.\ \ref{fig:Hex}. 

The honeycomb lattice coincides with one of the three Platonic (regular) tilings, namely the hexagonal tiling, whereas the other two Platonic 
tilings are the trivial triangular and square ones which are obtained from the triangular and square lattice by connecting the nearest neighbors to 
each other to constitute the prototiles. The structure on the right hand side in the lower panel of Fig.\ \ref{fig:Hex} 
corresponds to one of the eight Archimedean tilings, namely to the so called \textit{snub trihexagonal tiling}.

\subsubsection{Octagon-based structures (Oct)}

\begin{figure}[h!]
 \centering
 \includegraphics[height=5.0cm]{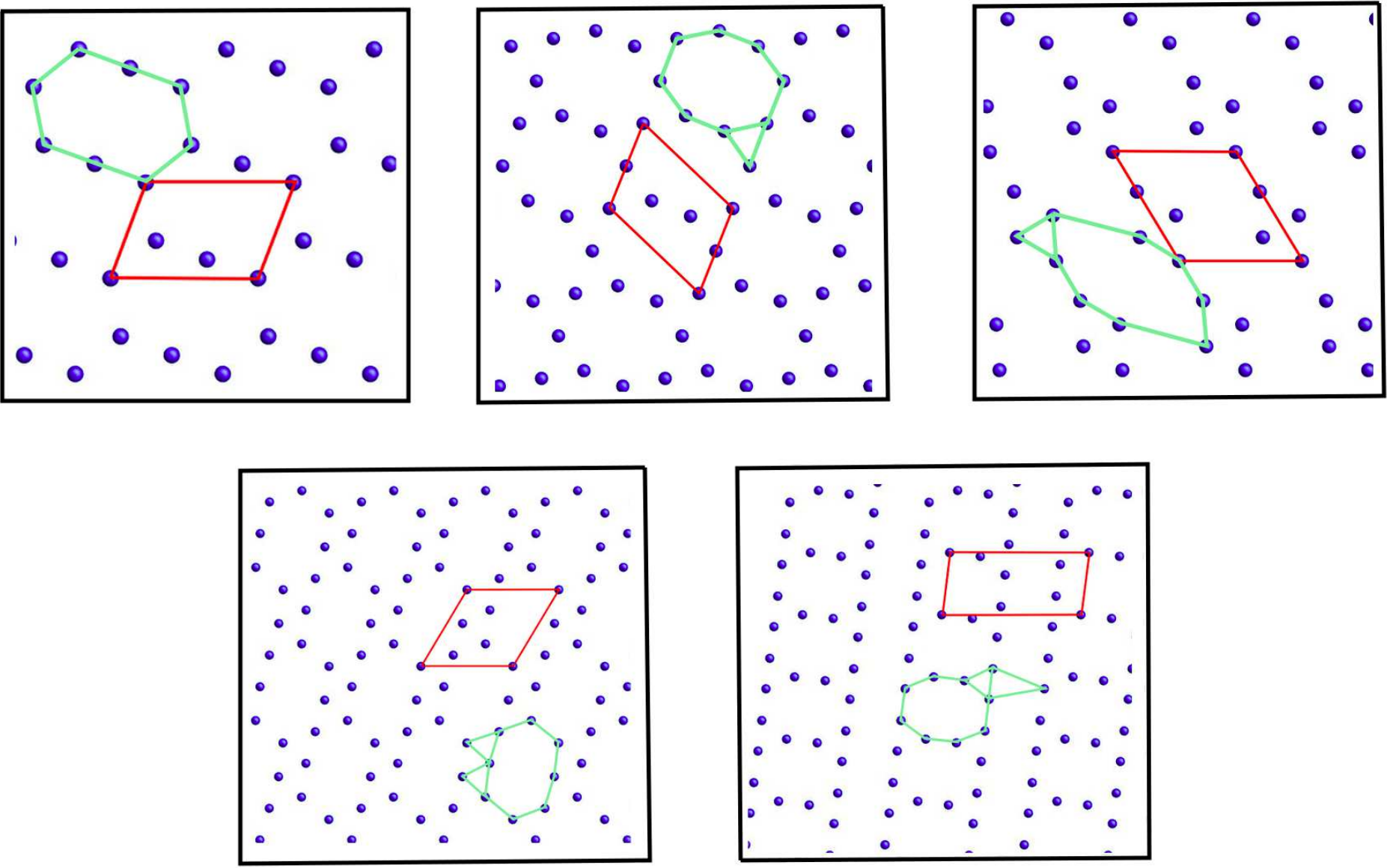}
 \caption{Schematic illustration of five stable octagon-based structures with non-regular octagons as their main repeating units
 	  at $T=0$ as obtained by lattice-sum minimizations. 
      	  The prototiles of the corresponding tilings are either one octagon or an octagon with one or two different triangles as indicated by green lines. 
	   Red lines mark the unit cells with $n=3,4,5$.}	     
 \label{fig:Oct}
\end{figure}

Octagon-based structures occur in the first non-triangular stability mode for $\kappa d > 1.5$ and $1.2 < \sqrt{\rho} d < 1.4$ 
and they are colored gray in Fig.\ \ref{fig:pd_detail} including the gray transparent region. We identify five different structures with a 
non-regular octagon as the main repeating unit, where four of them tile the space together with one or two different triangles (not necessarily equilateral). 
Hence, the prototiles are an octagon and one or two triangles as indicated by green lines in Fig.\ \ref{fig:Oct}. 
The red lines show a minimal unit cell for each structure.  

\subsubsection{Trimers (Tri)}

\begin{figure}[h!]
 \centering
 \includegraphics[height=3.0cm]{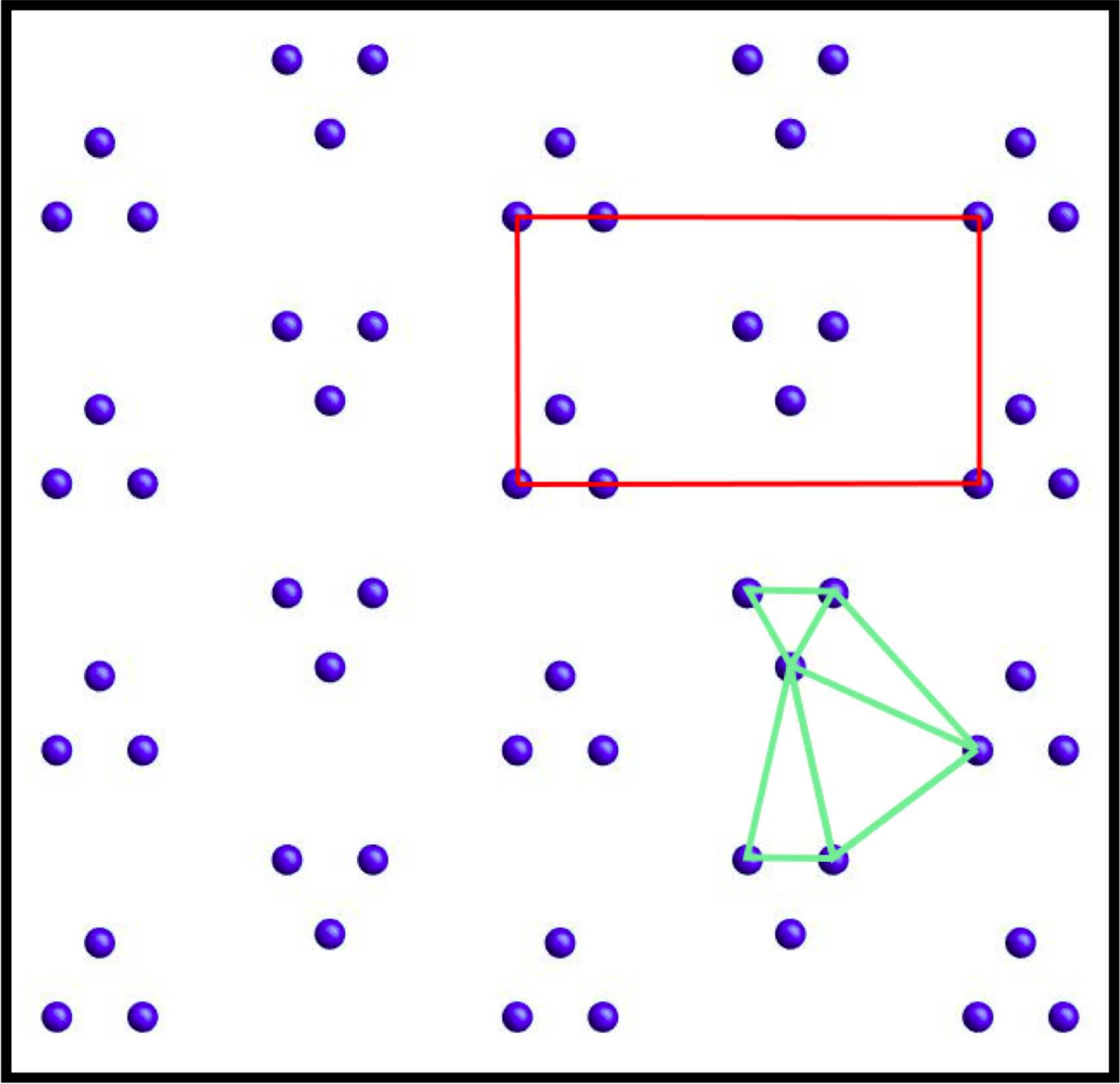}
 \caption{Schematic illustration of a stable trimer structure at $T=0$ as obtained by lattice-sum minimizations. 
 	  The corresponding tiling can be constructed by four different triangles as its prototiles which 
          are shown by green lines. Red lines mark the unit cell with $n=6$.}
 \label{fig:Tri}
\end{figure}

The trimer phase is found at $\kappa d > 1.7$ and $\sqrt{\rho} d \approx 0.9$ with the corresponding phase structures involving two well-separated
length scales per dimension. The larger one dictates the periodicity of the overall lattice, whereas at the smaller length scale, the particles
are arranged in equilateral triangles as a basis, see Fig.\ \ref{fig:Tri}. The corresponding tiling possesses four distinct triangles as indicated by the 
green lines in the same figure, where also a unit cell with $n=6$ is demonstrated by red lines.

\subsubsection{Triangle-rectangular structures (TR)}

\begin{figure}[h!]
 \centering
 \includegraphics[height=5.0cm]{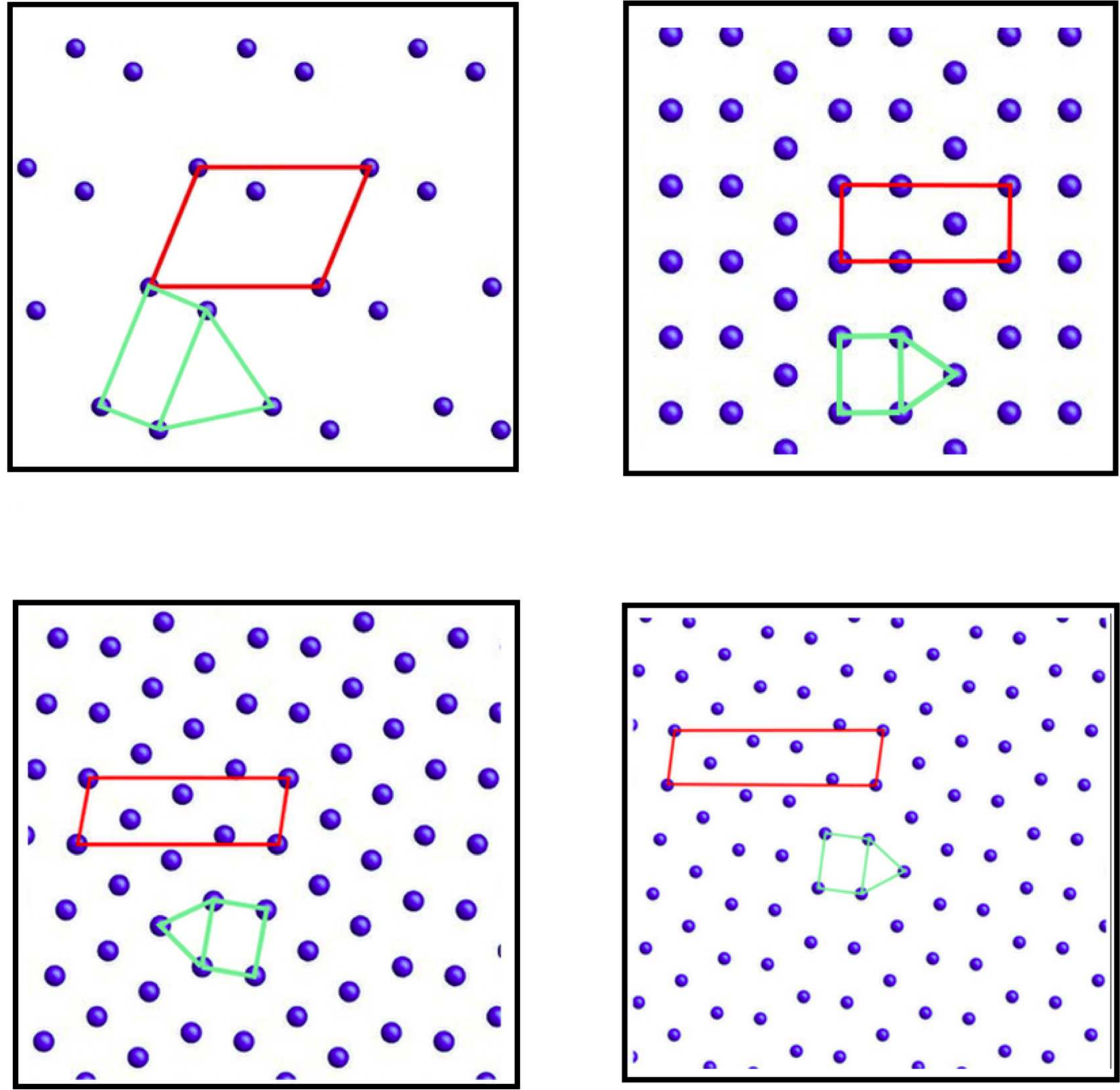}
 \caption{Schematic illustration of four stable triangle-rectangular structures at $T=0$ as obtained by lattice-sum minimizations. 
     	  The corresponding tiling has a rectangle and a triangle as its prototiles, whereas the ratio between them differs in each structure.
  	  Red lines mark the unit cells with $n=2,\cdots,5$.}
 \label{fig:TR}
\end{figure}

Here, we reveal four different types of lattice structures, where we choose to demonstrate the lattice points as arranged in triangles and rectangles 
as basic constituents (prototiles of the corresponding tilings, cf.\ green lines in Fig.\ \ref{fig:TR}).  
Each structure possesses a different ratio between these constituents. The upper left structure of Fig.\ \ref{fig:TR} occurs in the first non-triangular stability
mode ($\sqrt{\rho} d < 1.1$), whereas the others are identified for $\kappa d > 1.3$ and $2.1 < \sqrt{\rho} d < 2.3$. The stability zones are shown by red areas in 
Fig.\ \ref{fig:pd_detail}. As usual, the red lines in Fig.\ \ref{fig:TR} depict the unit cells with $n=2,\cdots,5$. 

\subsubsection{Triangle-square structures (TS)}

\begin{figure}[h!]
 \centering
 \includegraphics[height=3.0cm]{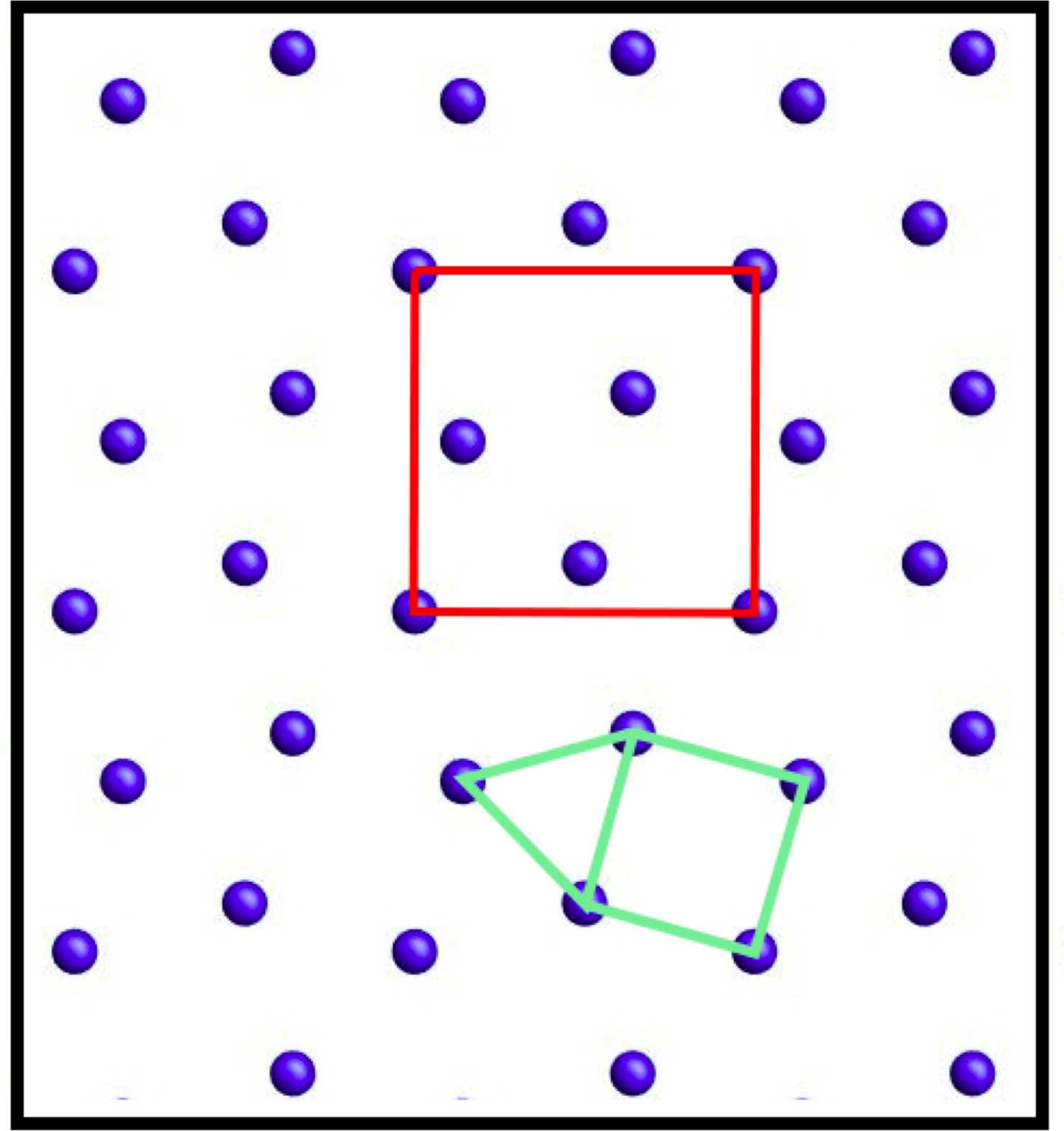}
 \caption{Schematic illustration of a stable triangle-square structure at $T=0$ as obtained by lattice-sum minimizations. It composes 
 	  of a square and a triangle as the prototiles of the corresponding tiling.
     	  Red lines mark the unit cell with $n=4$.}
 \label{fig:TS}
\end{figure}

Triangle-square phase is indicated by the turquoise region in the phase diagram in Fig.\ \ref{fig:pd_detail} with the phase structure possessing 
four particles in the unit cell (cf.\ red lines in Fig.\ \ref{fig:TS}). The corresponding tiling consisting of equilateral triangles and squares as specified by
the green lines in Fig.\ \ref{fig:TS} is revealed to be the \textit{snub square} tiling, also known as $\sigma$- or H-phase, which corresponds to one of the eight Archimedean tilings.   

\subsubsection{Pentagon-based structures (Pen)}

\begin{figure}[h!]
 \centering
 \includegraphics[height=10.0cm]{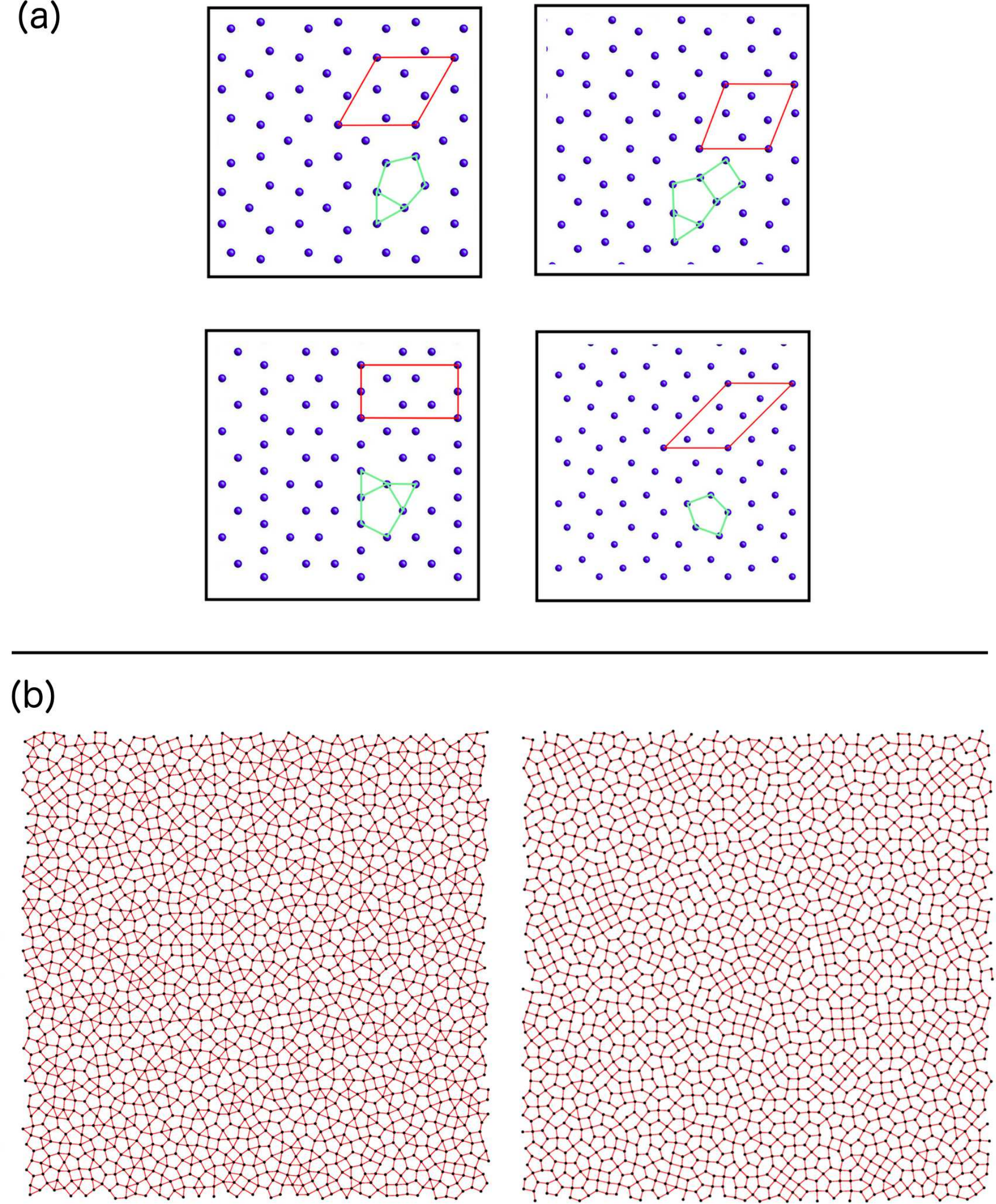}
 \caption{Stable pentagon-based structures at (a) $T=0$ as obtained by lattice-sum minimizations and (b) $T>0$ with non-regular pentagons 
          being the main repeating unit as obtained by Brownian Dynamics simulations. 
      	  The four crystalline ground-state structures are schematically illustrated in (a), where green lines indicate the different prototiles of the 
	  corresponding tilings, and red lines emphasize the unit cell in each structure. In (b), two characteristic simulation snapshots are shown exhibiting
          high local pentagonal order and resembling some of the pentagon-based structures from (a).}
 \label{fig:Pen3}
\end{figure}

In this group, we have four perfect ground-state lattice structures possessing $n=5$ ($n=6$) as shown in the upper (lower) level of Fig.\ \ref{fig:Pen3}a,  
with the red lines indicating the corresponding unit cells. These structures are found in a relatively broad regime in
the first non-triangular stability mode for $1.1 < \sqrt{\rho}d < 1.4$ around $\kappa \approx 1$ as well as in a tiny regime for
$\sqrt{\rho}d \approx 2.15$ and $\kappa d >1.7$. These regions are indicated by the purple transparent areas in Fig.\ \ref{fig:pd_detail}.

As seen in Fig.\ \ref{fig:Pen3}a, the pentagon-based structures differ in the type and number of prototiles, that is, some of them tile
the space periodically by just one non-regular pentagon, whereas others need (beside the pentagon) one or two different triangles or a triangle and a square.

In Fig.\ \ref{fig:Pen3}b, we demonstrate two characteristic snapshots of finite-temperature simulations showing clear local pentagonal orderings.
Black dots represent particle positions, whereas red lines have been introduced to connect the nearest neighbors based on a distance criterion, and 
thus to highlight a possible tessellation of the space with pentagons. The basic difference between both snapshots in Fig.\ \ref{fig:Pen3}b is that
the left one comprises predominantly pentagons and rhombuses as tiles, whereas in the right image there exists a considerable amount of triangles beside 
the pentagons, suggesting a strong resemblance to the ground-state structure shown on the upper left in Fig.\ \ref{fig:Pen3}a. 
We would like to mention that we have undertaken different runs starting from random, triangular, and square lattice configurations, 
and all runs lead to very similar final configurations displaying the same local pentagonal orderings at prescribed system parameters.

Our lattice-minimization routine reveals solely the stability of perfect lattices at zero-temperature. Having the possibility of
stable non-periodic structures in mind, we have included some quasicrystalline orderings according to the Penrose-, square-triangle- and 
square-rhombic-tiling into our calculations by computing the potential energies per particle of large periodic approximants
of the corresponding quasiperiodic tilings. As a result, within the studied parameter range, we have not observed any stable quasicrystalline phase.

\section{Conclusions}

In summary, we have developed a theoretical tool to predict the self-assembly of complex phase structures in classical condensed matter systems 
where thermodynamic stability is dictated by pair interactions. Our theory involves the enthalpy-like pair potential. We conjecture that complex  
non-trivial orderings  can occur if the enthalpy-like potential rather than the pair interaction itself possesses a concave region.
In order to validate our theory, we studied the specific example of two-dimensional colloid-polymer mixtures, which we have modeled with effective pair 
interactions involving a short-ranged depletion attraction and a long-ranged screened electrostatic Coulomb repulsion. 
In particular, we have investigated a detailed zero-temperature phase diagram of our system and we have analyzed the validity of occurring phases at finite 
but relatively small temperatures $T>0$. We found a remarkable agreement between our theory and ground-state calculations as we revealed that
our theory can identify relatively good the parameter region where complex structures with two or more length scales are stable. 

Furthermore, our ground-state phase diagram exhibits a rich morphology: First of all, we identify large regions of triangular-lattice stability, and regions  
of non-triangular lattices appearing as stability modes as a function of the density at fixed depletion length and polymer concentration. 
Second of all, the non-triangular regimes themselves feature a large diversity with respect to the stable phases. We recover simple phases such as rhombic and
square lattices, but more interestingly, we also reveal complex phase structures with pentagon-, hexagon-, and octagon-based structures as well as trimers, 
triangle-rectangular and triangle-square crystals, some of which are also found to be stable at $T>0$, and some others not. Furthermore, some of these 
ground-state structures correspond to the well-known Archimedean tilings, self-assembly of which has attracted a special interest in fundamental 
and applied physical sciences \cite{Sch10,Ant11,Mil14}. 

Ground state calculations can always depend on the candidate structures that are considered. We have determined all phases that are obtained for 
up to $n=6$ particles per unit cell, where we show the phases for up to $n=4$ particles per unit cell and indicate how an inclusion of $n=5,6$ particles 
into our minimization process yields slight changes shown by the transparent regions encircled by the red lines. The overall change when increasing 
the number of particles per unit cell is marginal and hence we only expect non-significant morphology changes of the non-triangular stability regions 
in the phase diagram in Fig.\ \ref{fig:pd_detail} upon an inclusion of $n>6$ particles per unit cell. 
Our Brownian Dynamics simulations confirm the triangular, square, rhombic, and pentagon-based structures. 
However, it is noteworthy that occurrence of exotic phases other than hitherto found ones within the stability zone 
of non-triangular lattices cannot be ultimately excluded. 

Our investigations have fundamental implications as we identified the ground-state phase diagram of colloid-polymer mixtures alongside a zoo of 
novel structures occurring in the same system, and practical implications as we establish systematic routes for the self-assembly of complex structures.
These phases should be accessible in experiments and therefore our phase diagram explains how to tailor complex colloidal structures that might, e.g., 
be of interest for photonic applications.

Finally, we want to stress that our theory involving the enthalpy-like pair potential can be applied to all particle interactions according to isotropic pair 
potentials and that it opens far-reaching possibilities to explore the self-assembly in numerous condensed matter phases.

\begin{acknowledgments}
We would like to thank A. S. Kraemer, F. Martelli, and R. E. Rozas for illuminating discussions. E.C.O. and M.S. received financial support from the
SFF of the HHU D{\"u}sseldorf. Furthermore, A.M. and M.S. were supported by the DFG within the Emmy Noether program (Grant Schm2657/2).
\end{acknowledgments}

\bibliographystyle{unsrt}
%\normalbaselines
%\bibliography{./collpol}

\begin{thebibliography}{10}

\bibitem{Whi91}
G.~M. Whitesides, J.~P. Mathias, and C.~T. Seto.
\newblock Molecular self-assembly and nanochemistry: a chemical strategy for
  the synthesis of nanostructures.
\newblock {\em Science}, 254:1312--1319, 1991.

\bibitem{Whi02}
G.~M. Whitesides and B.~Grzybowski.
\newblock Self-assemly at all scales.
\newblock {\em Science}, 295:2418, 2002.

\bibitem{Whi02_pnas}
G.~M. Whitesides and M.~Boncheva.
\newblock Beyond molecules: Self-assembly of mesoscopic and macroscopic
  components.
\newblock {\em Proc. Nat. Acad. Sci.}, 99:4769--4774, 2002.

\bibitem{Gro08}
R.~Gro{\ss} and M.~Dorigo.
\newblock Self-assembly at the macroscopic scale.
\newblock {\em Proc. IEEE}, 96:1490--1508, 2008.

\bibitem{Leh02}
J.-M. Lehn.
\newblock Toward self-organization and complex matter.
\newblock {\em Science}, 295:2400--2403, 2002.

\bibitem{Jen99}
S.~A. Jenekhe and X.~L. Chen.
\newblock Self-assembly of ordered microporous materials from rod-coil block
  copolymers.
\newblock {\em Science}, 283:372, 1999.

\bibitem{Jac04}
A.~M. Jackson, J.~W. Myerson, and F.~Stellacci.
\newblock Spontaneous assembly of subnanometre-ordered domains in the ligand
  shell of monolayer-protected nanoparticles.
\newblock {\em Nat. Mater.}, 3:330, 2004.

\bibitem{Man03}
V.~N. Manoharan, M.~T. Elsesser, and D.~J. Pine.
\newblock Dense packing and symmetry in small clusters of microspheres.
\newblock {\em Science}, 301:483, 2003.

\bibitem{Jac16}
W.~M. Jacobs and D.~Frenkel.
\newblock Self-assembly of structures with addressable complexity.
\newblock {\em J. Am. Chem. Soc.}, 138:2457--2467, 2016.

\bibitem{Kus69}
D.~J. Kushner.
\newblock Self-assembly of biological structures.
\newblock {\em Bacteriol Rev.}, 33:302--345, 1969.

\bibitem{Yve94}
Y.~Engelborghs.
\newblock Microtubules: dissipative structures formed by self-assembly.
\newblock {\em Biosensors \& Bioelectronics}, 9:685--689, 1994.

\bibitem{Bud95}
E.~O. Budrene and H.~C. Berg.
\newblock Dynamics of formation of symmetrical patterns by chemotactic
  bacteria.
\newblock {\em Nature}, 376:49--53, 1995.

\bibitem{Jul08}
J.~D. Halley and D.~A. Winkler.
\newblock Critical-like self-organization and natural selection: Two facets of
  a single evolutionary process?
\newblock {\em BioSystems}, 92:148--158, 2008.

\bibitem{Dot14}
T.~Dotera, T.~Oshiro, and P.~Ziherl.
\newblock Mosaic two-lengthscale quasicrystals.
\newblock {\em Nature}, 506:208--211, 2014.

\bibitem{She84}
D.~Shechtman, I.~Blech, D.~Gratias, and J.~W. Cahn.
\newblock Metallic phase with long range orientational order and no translation
  symmetry.
\newblock {\em Phys. Rev. Lett.}, 53:1951--1954, 1984.

\bibitem{Ass07}
L.~Assoud, R.~Messina, and H.~L{\"o}wen.
\newblock Stable crystalline lattices in two-dimensional binary mixtures.
\newblock {\em Eur. Phys. Lett.}, 80:48001, 2007.

\bibitem{Ass08}
L.~Assoud, R.~Messina, and H.~L{\"o}wen.
\newblock Binary crystals in two-dimensional two-component yukawa mixtures.
\newblock {\em J. Chem. Phys.}, 129:164511, 2008.

\bibitem{Ant13}
M.~Antlanger and G.~Kahl.
\newblock Wigner crystals for a planar, equimolar binary mixture of classical,
  charged particles.
\newblock {\em Condens. Matter Phys.}, 16:43501, 2013.

\bibitem{Dam12}
P.~F. Damasceno, M.~Engel, and S.~C. Glotzer.
\newblock Predictive self-assembly of polyhedra into complex structures.
\newblock {\em Science}, 337:453--457, 2012.

\bibitem{Che11}
Q.~Chen, S.~C. Bae, and S.~Granick.
\newblock Directed self-assembly of a colloidal kagome lattice.
\newblock {\em Nature}, 469:381, 2011.

\bibitem{Dop12}
G.~Doppelbauer, E.~G. Noya, E.~Bianchi, and G.~Kahl.
\newblock Self-assembly scenarios of patchy colloidal particles.
\newblock {\em Soft Matter}, 8:7768, 2012.

\bibitem{Bia14}
E.~Bianchi, C.~N. Likos, and G.~Kahl.
\newblock Tunable assembly of heterogeneously charged colloids.
\newblock {\em Nano Lett.}, 14:3412--3418, 2014.

\bibitem{Rec05}
M.~Rechtsman, F.~H. Stillinger, and S.~Torquato.
\newblock Optimized interactions for targeted self-assembly: Application to a
  honeycomb lattice.
\newblock {\em Phys. Rev. Lett.}, 95:228301, 2005.

\bibitem{Rec06}
M.~Rechtsman, F.~H. Stillinger, and S.~Torquato.
\newblock Designed interaction potentials via inverse methods for
  self-assembly.
\newblock {\em Phys. Rev. E}, 73:011406, 2006.

\bibitem{Bat08}
R.~D. Batten, F.~H. Stillinger, and S.~Torquato.
\newblock Classical disordered ground states: Super-ideal gases and stealth and
  equi-luminous materials.
\newblock {\em J. Appl. Phys.}, 104:033504, 2008.

\bibitem{reviewSal}
S.~Torquato.
\newblock Inverse optimization techniques for targeted self-assembly.
\newblock {\em Soft Matter}, 5:1157--1173, 2009.

\bibitem{Coh09}
H.~Cohn and A.~Kumar.
\newblock Algorithmic design of self-assembling structures.
\newblock {\em Proc. Nat. Acad. Sci.}, 106:9570--9575, 2009.

\bibitem{Mar11}
{\'E}.~Marcotte, F.~H. Stillinger, and S.~Torquato.
\newblock Unusual ground states via monotonic convex pair potentials.
\newblock {\em J. Chem. Phys.}, 164105:134, 2011.

\bibitem{Jain13}
A.~Jain, J.~R. Errington, and T.~M. Truskett.
\newblock Inverse design of simple pairwise interactions with low coordinated
  3d lattice ground states.
\newblock {\em Soft Matter}, 9:3866, 2013.

\bibitem{Jain14}
A.~Jain, J.~R. Errington, and T.~M. Truskett.
\newblock Dimensionality and design of isotropic interactions that stabilize
  honeycomb, square, simple cubic, and diamond lattices.
\newblock {\em Phys. Rev. X}, 4:031049, 2014.

\bibitem{Pin16a}
W.~D. Pi{\~n}eros, M.~Baldea, and T.~M. Truskett.
\newblock Designing convex repulsive pair potentials that favor assembly of
  kagome and snub square lattices.
\newblock {\em J. Chem. Phys.}, 145:054901, 2016.

\bibitem{Pin16b}
W.~D. Pi{\~n}eros, M.~Baldea, and T.~M. Truskett.
\newblock Breadth versus depth: Interactions that stabilize particle assemblies
  to changes in density or temperature.
\newblock {\em J. Chem. Phys.}, 144:084502, 2016.

\bibitem{Cas03}
R.~Casta{\~n}eda-Priego, A.~Rodr{\'i}guez-L{\'o}pez, and J.~M.
  M{\'e}ndez-Alcaraz.
\newblock Depletion forces in two-dimensional colloidal mixtures.
\newblock {\em J. Phys.: Condens. Matter}, 15:S3393--S3409, 2003.

\bibitem{Dom12}
P.~Dom{\'i}nguez-Garc{\'i}a.
\newblock Microrheological consequences of attractive colloid-colloid
  potentials in a two-dimensional brownian fluid.
\newblock {\em Eur. Phys. J. E}, 35:73, 2012.

\bibitem{Fen15}
L.~Feng, B.~Laderman, S.~Sacanna, and P.~Chaikin.
\newblock Re--entrant solidification in polymer-–colloid mixtures as a
  consequence of competing entropic and enthalpic attractions.
\newblock {\em Nat. Mater.}, 14:61--65, 2015.

\bibitem{Pha02}
K.~N. Pham, A.~M. Puertas, J.~Bergenholtz, S.~U. Egelhaaf, A.~Moussaid, P.~N.
  Pusey, A.~B. Schofield, M.~E. Cates, M.~Fuchs, and W.~C.~K. Poon.
\newblock Multiple glassy states in a simple model system.
\newblock {\em Science}, 296:104--106, 2002.

\bibitem{Eck02}
T.~Eckert and E.~Bartsch.
\newblock Re-entrant glass transition in a colloid-polymer mixture with
  depletion attractions.
\newblock {\em Phys. Rev. Lett.}, 89:125701, 2002.

\bibitem{Man05}
S.~Manley, H.~M. Wyss, K.~Miyazaki, J.~C. Conrad, V.~Trappe, L.~J. Kaufman,
  D.~R. Reichman, and D.~A. Weitz.
\newblock Glasslike arrest in spinodal decomposition as a route to colloidal
  gelation.
\newblock {\em Phys. Rev. Lett.}, 95:238302, 2005.

\bibitem{Lu08}
P.~J. Lu, E.~Zaccarelli, F.~Ciulla, A.~B. Schofield, F.~Sciortino, and D.~A.
  Weitz.
\newblock Gelation of particles with short--range attraction.
\newblock {\em Nature}, 453:499--504, 2008.

\bibitem{Zha13}
I.~Zhang, C.~P. Royall, M.~A. Faers, and P.~Bartlett.
\newblock Phase separation dynamics in colloid--polymer mixtures: the effect of
  interaction range.
\newblock {\em Soft MAtter}, 9:2076--2084, 2013.

\bibitem{Man14}
Ethayaraja Mani, Wolfgang Lechner, Willem~K. Kegel, and Peter~G. Bolhuis.
\newblock Equilibrium and non-equilibrium cluster phases in colloids with
  competing interactions.
\newblock {\em Soft Matter}, 10:4479--4486, 2014.

\bibitem{Koh16}
M.~Kohl, R.~F. Capellmann, M.~Laurati, S.~U. Egelhaaf, and M.~Schmiedeberg.
\newblock Directed percolation identified as equilibrium pre--transition
  towards non-equilibrium arrested gel states.
\newblock {\em Nat. Commun.}, 7:11817, 2016.

\bibitem{Str04}
A.~Stradner, H.~Sedgwick, F.~Cardinaux, W.~C.~K. Poon, S.~U. Egelhaaf, and
  P.~Schurtenberger.
\newblock Equilibrium cluster formation in concentrated protein solutions and
  colloids.
\newblock {\em Nature}, 432:492--495, 2004.

\bibitem{Taf10}
J.~Taffs, A.~Malins, S.~R. Williams, and C.~P. Royall.
\newblock A structural comparison of models of colloid--polymer mixtures.
\newblock {\em J. Phys.: Condens. Matter}, 22:104119, 2010.

\bibitem{Cal93}
F.~Leal Calderon, J.~Bibette, and J.~Biais.
\newblock Experimental phase diagrams of polymer and colloid mixtures.
\newblock {\em EPL (Europhysics Letters)}, 23(9):653, 1993.

\bibitem{Dij99}
M.~Dijkstra, R.~van Roij, and R.~Evans.
\newblock Phase diagram of highly aymmetric binary hard-shpere mixtures.
\newblock {\em Phys. Rev. E}, 59:5744--5771, 1999.

\bibitem{Sch02}
M.~Schmidt, H.~L{\"o}wen, J.~M. Brader, and R.~Evans.
\newblock Density functional theory for a model colloid--polymer mixture: bulk
  fluid phases.
\newblock {\em J. Phys.: Condens. Matter}, 14:9353--9382, 2002.

\bibitem{Poo02}
W.~C.~K. Poon.
\newblock The physics of a model colloid--polymer mixture.
\newblock {\em J. Phys.: Condens. Matter}, 14:R859--R880, 2002.

\bibitem{Aar02}
D~G A~L Aarts, R~Tuinier, and H~N~W Lekkerkerker.
\newblock Phase behaviour of mixtures of colloidal spheres and excluded-volume
  polymer chains.
\newblock {\em Journal of Physics: Condensed Matter}, 14(33):7551, 2002.

\bibitem{Roy05}
C.~P. Royall, D.~G. A.~L. Aarts, and H.~Tanaka.
\newblock Fluid structure in colloid-polymer mixtures: the competition between
  electrostatics and depletion.
\newblock {\em J. Phys.: Condens. Matter}, 17:S3401--S3408, 2005.

\bibitem{For05}
A.~Fortini, M.~Dijkstra, and R.~Tuinier.
\newblock Phase behaviour of charged colloidal sphere dispersions with added
  polymer chains.
\newblock {\em J. Phys.: Condens. Matter}, 17:7783--7803, 2005.

\bibitem{Fle07}
Gerard~J. Fleer and Remco Tuinier.
\newblock Analytical phase diagram for colloid-polymer mixtures.
\newblock {\em Phys. Rev. E}, 76:041802, Oct 2007.

\bibitem{Fle08}
Gerard~J. Fleer and Remco Tuinier.
\newblock Analytical phase diagrams for colloids and non-adsorbing polymer.
\newblock {\em Advances in Colloid and Interface Science}, 143(1–2):1 -- 47,
  2008.

\bibitem{Tui08}
R.~Tuinier, P.~A. Smith, W.~C.~K. Poon, S.~U. Egelhaaf, D.~G. A.~L. Aarts,
  H.~N.~W. Lekkerkerker, and G.~J. Fleer.
\newblock Phase diagram for a mixture of colloids and polymers with equal size.
\newblock {\em EPL (Europhysics Letters)}, 82(6):68002, 2008.

\bibitem{Gar16}
\'Alvaro Gonz\'alez~Garc\'{\i}a and Remco Tuinier.
\newblock Tuning the phase diagram of colloid-polymer mixtures via yukawa
  interactions.
\newblock {\em Phys. Rev. E}, 94:062607, Dec 2016.

\bibitem{Jor03}
A.~Moncho-Jord{\'a}, A.~A. Loius, P.~G. Bolhuis, and R.~Roth.
\newblock The asakura--oosawa model in the protein limit: the role of
  many--body interactions.
\newblock {\em J. Phys.: Condens. Matter}, 15:S3429--S3442, 2003.

\bibitem{Dij06}
M.~Dijkstra, R.~van Roij, R.~Roth, and A.~Fortini.
\newblock Effect of many--body interactions on the bulk and interfacial phase
  behavior of a model colloid-polymer mixture.
\newblock {\em Phys. Rev. E}, 73:041404, 2006.

\bibitem{Cui05}
B.~Cui, B.~Lin, D.~Frydel, and S.~A. Rice.
\newblock Anomalous behavior of the depletion potential in
  quasi-two-dimensional binary mixtures.
\newblock {\em Phys. Rev. E}, 72:021402, 2005.

\bibitem{Fry05}
D.~Frydel and S.~A. Rice.
\newblock Depletion interaction in a quasi-two-dimensional colloid assembly.
\newblock {\em Phys. Rev. E}, 71:041402, 2005.

\bibitem{For06}
A.~Fortini, M.~Schmidt, and M.~Dijkstra.
\newblock Phase behavior and structure of model colloid-polymer mixtures
  confined between two parallel planar walls.
\newblock {\em Phys. Rev. E}, 73:051502, 2006.

\bibitem{Vin06}
R.~L.~C. Vink, K.~Binder, and J.~Horbach.
\newblock Critical behavior of a colloid-polymer mixture confined between
  walls.
\newblock {\em Phys. Rev. E}, 73:056118, 2006.

\bibitem{Kepler1619}
J.~Kepler.
\newblock {\em Harmonices Mundi}.
\newblock 1619.

\bibitem{Sch10}
M.~Schmiedeberg, J.~Mikhael, S.~Rausch, J.~Roth, L.~Helden, C.~Bechinger, and
  H.~Stark.
\newblock Archimedean-like colloidal tilings on substrates with decagonal and
  tetradecagonal symmetry.
\newblock {\em Eur. Phys. J. E}, 32:25--34, 2010.

\bibitem{Ant11}
M.~Antlanger, G.~Doppelbauer, and G.~Kahl.
\newblock On the stability of archimedean tilings formed by patchy particles.
\newblock {\em J. Phys.: Condens. Matter}, 23:404206, 2011.

\bibitem{Mil14}
J.~A. Millan, D.~Ortiz, G.~van Anders, and S.~C. Glotzer.
\newblock Self-assembly of archimedean tilings with enthalpically and
  entropically patchy polygons.
\newblock {\em ACS Nano}, 8:2918--2928, 2014.

\bibitem{Asa54}
S.~Asakura and F.~Oosawa.
\newblock {\em J. Chem. Phys.}, 22:1255, 1954.

\bibitem{Vri76}
A.~Vrij.
\newblock {\em Pure Appl. Chem.}, 48:471, 1976.

\bibitem{Nel65}
J~.~A. Nelder and R.~A. Mead.
\newblock A simplex method for function minimization.
\newblock {\em Computer Journal}, 7:308--313, 1965.

\bibitem{Slo87}
N.~J.~A. Sloane.
\newblock Theta series and magic numbers for diamond and certain ionic crystal
  structures.
\newblock {\em J. Math. Phys.}, 28:1653, 1987.

\bibitem{Con99}
J.~H. Conway and N.~J.~A. Sloane.
\newblock {\em Sphere Packings, Lattices and Groups}.
\newblock Springer, New York, 3rd edn. edition, 1999.

\bibitem{Gra12}
J.~de~Graaf, L.~Filion, M.~Marechal, R.~van Roij, and M.~Dijkstra.
\newblock Crystal-structure prediction via the floppy-box monte carlo
  algorithm: Method and application to hard (non)convex particles.
\newblock {\em J. Chem. Phys.}, 137:214101, 2012.

\end{thebibliography}

%\newpage
%\appendix
%%%%%%%%%%
%\section{Calculation of $S(k)$ for Fibonacci projection tilings} 
%\label{app:Sk}
%%%%%%%%%%

\end{document}